\documentclass[12pt,draftcls,onecolumn]{IEEEtran}

\makeatletter
\def\ps@headings{%
\def\@oddhead{\mbox{}\scriptsize\rightmark \hfil \thepage}%
\def\@evenhead{\scriptsize\thepage \hfil \leftmark\mbox{}}%
\def\@oddfoot{}%
\def\@evenfoot{}}
\makeatother

\usepackage{amstext,amsmath,amssymb}
\usepackage{times}
\usepackage{graphicx}
\usepackage{algorithmic}
\usepackage{algorithm}
\usepackage{url}
\usepackage{graphicx}
\usepackage{color}
\usepackage{verbatim}
\usepackage{cite}
\usepackage{subfig}

\usepackage{multirow}

\newtheorem{theorem}{Theorem}
\newtheorem{lemma}[theorem]{Lemma}
\newtheorem{corollary}[theorem]{Corollary}

\newtheorem {fact}[theorem]{Fact}

\newcommand{\bdm}{
    \begin{displaymath}}

\newcommand{\edm}{
    \end{displaymath}}

\newcommand{\be}{
    \begin{equation}}

\newcommand{\ee}{
    \end{equation}}

\newcommand{\bea}{
    \begin{eqnarray}}

\newcommand{\eea}{
    \end{eqnarray}}

\newcommand{\beit}{\begin{itemize}}
\newcommand{\eeit}{\end{itemize}}

\newcommand{\eat}[1]{ }

\renewcommand{\algorithmiccomment}[1]{// {\em #1}}

\begin{document}
%
\title{Content Distribution by Multiple Multicast
Trees and Intersession Cooperation: Optimal Algorithms and Approximations}

\author{Xiaoying~Zheng,
        Chunglae~Cho,
        and~Ye~Xia
\thanks{X. Zheng is with Shanghai Advanced Research Institute, Chinese Academy of Sciences, and
Shanghai Research Center for Wireless Communications, China.
E-mail: zhengxy@sari.ac.cn.}%
\thanks{C. Cho and Y. Xia are with the Department of Computer and Information Science and Engineering, University of Florida, Gainesville, FL 32611. E-mail: \{ccho, yx1\}@cise.ufl.edu.}
}

\maketitle

\begin{abstract}

In traditional massive content distribution with multiple sessions, the sessions form separate overlay networks and operate independently, where some sessions may suffer from insufficient resources even though other sessions have excessive resources. To cope with this problem, we consider the universal swarming approach, which allows multiple sessions to cooperate with each other. We formulate the problem of finding the optimal resource allocation to maximize the sum of the session utilities and present a subgradient algorithm which converges to the optimal solution in the time-average sense. The solution involves an NP-hard subproblem of finding a minimum-cost Steiner tree. We cope with this difficulty by using a column generation method, which reduces the number of Steiner-tree computations. Furthermore, we allow the use of approximate solutions to the Steiner-tree subproblem. We show that the approximation ratio to the overall problem turns out to be no less than
the reciprocal of the approximation ratio to the Steiner-tree subproblem. Simulation results demonstrate that universal swarming improves the performance of resource-poor sessions with negligible impact to resource-rich sessions. The proposed approach and algorithm are expected to be useful for infrastructure-based content distribution networks with long-lasting sessions and relatively stable network environment.

\end{abstract}

\begin{keywords}
Multi-Tree Multicast, Optimization, Rate Allocation, Tree Packing,
Universal Swarming, Column Generation, Subgradient Algorithm
\end{keywords}


\section{Introduction}
\label{section:introduction}

The Internet is being applied to transfer content on a more and
more massive scale. While many content distribution techniques
have been introduced, most of the recently introductions are based
on the \emph{swarming} technique, such as FastReplica
\cite{LeeV07}, Bullet \cite{Bullet, Bullet'}, Chunkcast
\cite{Chunkcast}, BitTorrent \cite{BitTorrent}, and CoBlitz
\cite{CoBlitz}. In a swarming session, the file to be distributed
is broken into many chunks at the source node, which are then
spread out to the receivers; the receivers will then help each
other with the retrieval of the missing chunks. By taking
advantage of the resources of the receivers, swarming dramatically
improves the distribution efficiency (e.g., average downloading
rate, completion time) compared to the traditional
client-server-based approach.

The swarming technique was originally created by the end-user
communities for peer-to-peer (P2P) file sharing.
The subject of
this paper is how to apply swarming to infrastructure-based
content distribution and make the distribution more efficient.
Compared with the dynamic end-user file-sharing situation,
such infrastructure networks are often of medium size, consisting of hundreds of nodes, dedicated, and have much more stable nodes and links. 
In this setting, we will see that it is beneficial to view
swarming as distribution over multiple multicast trees. This view
allows us to pose the question of how to optimally distribute the
content (see \cite{ZCXinfocom08}).


The specific problem addressed in this paper is how to conduct
content distribution more efficiently in a network where multiple
distribution sessions coexist. A distribution {\em session}
consists of a file to be distributed, one or more sources and all
the nodes who wish to receive the file, i.e., the receivers.
Different sessions may have heterogenous resource capacities, such
as the source upload bandwidth, receiver download bandwidth, or
aggregate upload bandwidth. For instance, there may exist some
sessions with excessive {\em aggregate} upload bandwidth because
their throughput bottleneck is at the source upload bandwidth, the
receiver download bandwidth, or the internal network; at the same
time, there may exist some other sessions whose throughput
bottleneck is at their aggregate upload bandwidth. In the
traditional swarming approach, the sessions operate independently
by each forming a separate overlay network; this will be called
\emph{separate swarming}, which does not provide the opportunity
for the resource-poor sessions to use the surplus resources of the
resource-rich sessions. However, if we conduct \emph{universal
swarming}, that is, we combine multiple sessions together into a
single ``super session'' on a shared overlay network and allow
them to share each other's resources, the distribution efficiency
of the resource-poor sessions can improve greatly with negligible
impact on the resource-rich sessions, provided the resource-rich sessions have sufficient surplus resources. The paper examines
algorithms and theoretical issues related to universal swarming.
In universal swarming, a
distribution tree not only includes all the receivers interested
in downloading the file but may also contain nodes that are not
interested in the file; the latter will be called {\em
out-of-session nodes},
and the source and receivers are called {\em in-session nodes}.
Thus, each distribution tree for a session
is a \emph{Steiner} tree rooted at the source covering
all the receivers and the out-of-session nodes on the
tree are Steiner nodes\footnote{Given a directed graph $G=(V,E)$, and a subset
$V^{\prime} \subseteq V$ of vertices, a Steiner tree is
a connected and acyclic subgraph of $G$ which spans
all vertices of $V^{\prime}$. More information can be found
in \cite{Charikar98, Zosin02, Hsieh06}.}.

\begin{figure} [htbp]
\begin{minipage}{2in}
\begin{center}
\includegraphics[width=1in]{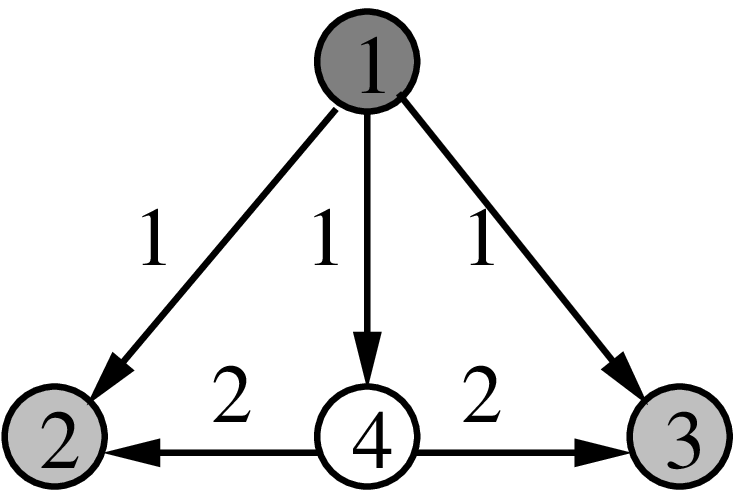}
\\ (a)
\end{center}
\end{minipage}
\begin{minipage}{4.2in}
\begin{center}
\includegraphics[width=2.5in]{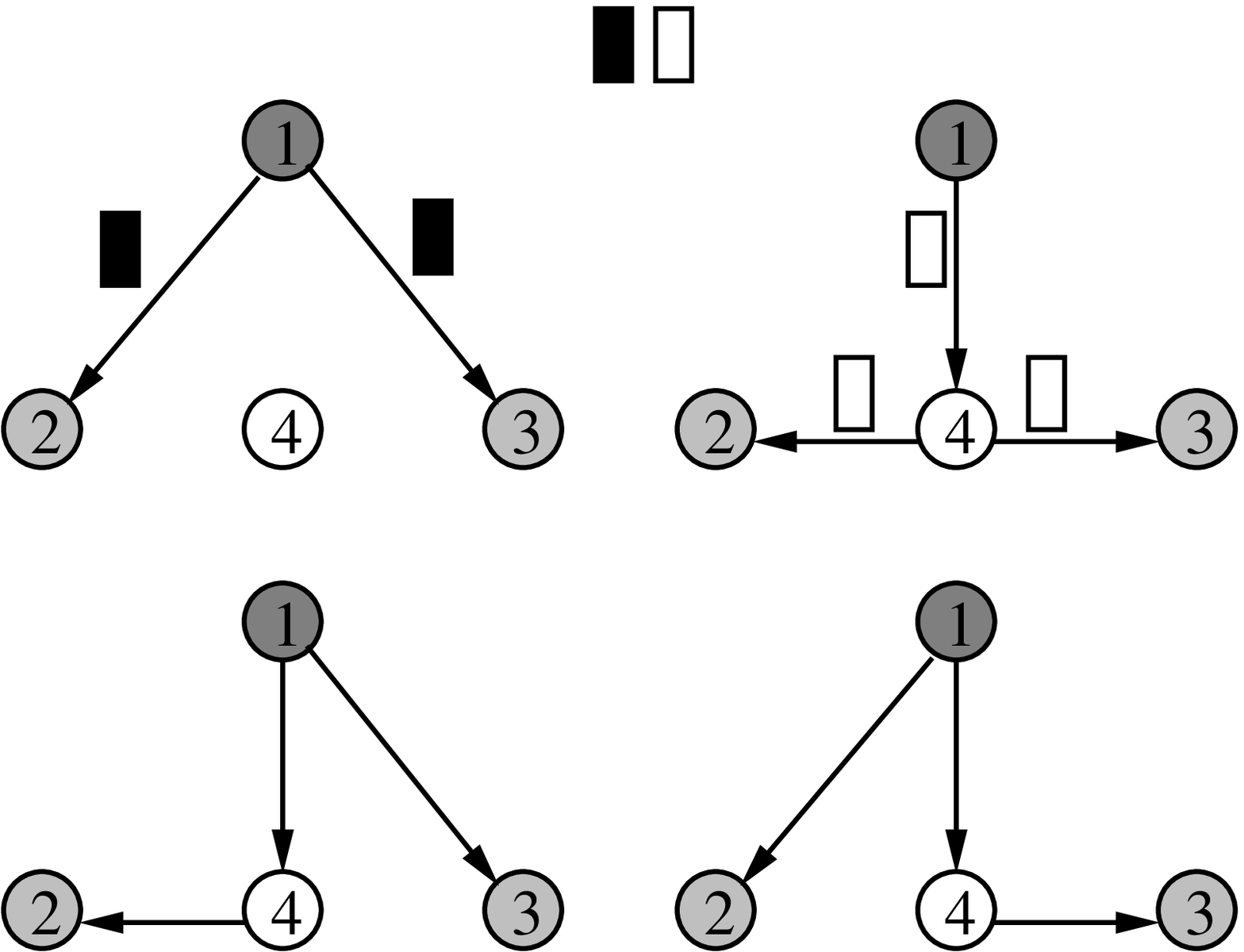}
\\ (b)
\end{center}
\end{minipage}
\vspace{-0.0in}
\caption{An universal swarming example. (a) Node 1 distributes a file to node 2 and 3; node 4 is an out-of-session node. The numbers
next to the links are the link capacities.
(b) All possible distribution trees and the optimal solution for throughput maximization. The boxes represent file chunks. Three of the trees involve the out-of-session node 4.}
\label{fig:dist_tree}
\end{figure}

To illustrate the main ideas, consider the toy example in
Fig. \ref{fig:dist_tree} (a). The numbers associated with the
links are their capacities. Node 1, 2 and 3 form a multicast session, and a large file is distributed from the source node 1 to the receivers, node 2 and 3. Node 4 is an out-of-session node. Suppose the file is split into many chunks at node 1. To distribute a chunk to the receivers, the chunk must travel down some tree rooted at the source and covering both receivers.
All possible distribution trees are shown in Fig. \ref{fig:dist_tree}
(b). Observe that, except the first tree, the other three trees include the out-of-session node. Fig. \ref{fig:dist_tree}
(b) shows an optimal rate allocation with respect to the objective of maximizing the total distribution rate, or equivalently, minimizing the required distribution time. The scenario is an example of universal swarming since the out-of-session node is used. For separate swarming, only the first tree can be used and the distribution rate is only one half as much.

\eat{
With the tree-based model, the optimal distribution problem can be
formulated as finding an optimal rate allocation on the multiple
multicast trees so that it achieves the optimal performance
objective. A version of this problem was addressed in
\cite{ZCXinfocom08} in the context of separate swarming.
The
rate-allocation problem in universal swarming, which this paper
concerns, is substantially more difficult. The main reason is
that, by the optimality condition, an optimal solution typically
uses only the minimum-cost trees to distribute the file chunks.
In \cite{ZCXinfocom08}, an overlay network consisting
of in-session nodes is constructed above the underlay IP
network for each multicast session. The topology of
each overlay network is pre-specified which makes
the separate swarming scheme suboptimal compared to
the universal swarming scheme.
It turns out that in the setting of separate
swarming, the algorithm only needs a minimum-cost (min-cost)
spanning tree since all the nodes in an overlay network
participate in transferring chunks.
However, an optimal universal swarming algorithm usually involves an
NP-hard subproblem of finding a min-cost Steiner
tree, and the approach proposed in \cite{ZCXinfocom08}
is unable to handle this difficulty. How to cope with this issue is one of the main themes in
this paper.
}

With the tree-based model, the optimal distribution problem can be
formulated as finding an optimal rate allocation on the
multicast trees so that it achieves the optimal performance
objective. A version of this problem was considered in
\cite{ZCXinfocom08} and its longer version \cite{ZhengCX10}
in the context of separate swarming.
The rate-allocation problem in universal swarming, which this paper
concerns, is substantially more difficult. The main reason is
that, by the optimality condition, an optimal solution typically
uses only the minimum-cost (min-cost) trees to distribute the file chunks.
In \cite{ZCXinfocom08}, for each multicast session, an overlay network consisting
of only in-session nodes is constructed above the underlay physical
network, where the topology of
each overlay network is pre-specified.
The algorithm for separate
swarming in \cite{ZCXinfocom08} only needs min-cost
{\em spanning} trees; finding a min-cost spanning tree is considered an easy problem since polynomial algorithms exist.
In contrast, an optimal universal swarming algorithm usually involves repeatedly
finding a min-cost {\em Steiner}
tree, which is an NP-hard problem. How to cope with this difficult issue is one of the main themes in
this paper. The approach proposed in \cite{ZCXinfocom08}
is unable to handle this difficulty.
On the positive side, since universal swarming corresponds to a less restricted way of doing multicast than separate swarming, performance improvement is expected. The degree of improvement can sometime be large.

We present two solution approaches, which can be used in
combination.
First, we incorporate into our rate-allocation
algorithm a column generation method, which can reduce the number
of times the min-cost Steiner tree is computed. Second, we allow
the use of approximate solutions to the Steiner-tree subproblem.
Such approximate solutions on directed
graphs can be found in \cite{Charikar98, Zosin02, Hsieh06}. When the above two methods are put together, the combined algorithm is rather difficult to analyze.
For the most part, there are little standard results that can be used directly to prove algorithm convergence or to give an approximation ratio to the rate-allocation problem when approximate min-cost trees are used in each iteration step.
One of our main technical contributions is to show the combined algorithm converges to solutions with performance guarantee. Specifically, the approximation ratio to the overall
rate-allocation problem is no less than the
reciprocal of the approximation ratio to the
Steiner-tree subproblem.

The overall rate-allocation algorithm that we will present is a
subgradient algorithm. It has the characteristic of assigning
positive rate to a single multicast tree per session at each
iteration; the rate assigned to the tree is computed based on the
link prices at the iteration. We can show that even though the
assigned rates in each iteration usually exceed the capacities of
some links, the time-average rates satisfy the link capacity
constraints, and eventually the rate allocation to each session
converges to the optimum (provided the Steiner-tree subproblem is
solved optimally.) It is worth pointing out that other
optimization algorithms may also be used here instead of the
subgradient algorithm.

The paper is organized as follows. The formal problem description
is given in Section \ref{section:models}. The subgradient
algorithm and its convergence proof are given in Section
\ref{sec:subgradient}. In Section \ref{sec:colgen}, we present the
column generation approach, combine it with the subgradient
algorithm, and study the performance bound when approximation
algorithms are applied to the minimum-cost tree subproblem. We
show some simulation results about our approach in Section
\ref{section:experiment}. In Section \ref{section:related},
we discuss additional related works.
The conclusion is drawn in Section
\ref{sec:conclusion}.

\section{Problem Description}
\label{section:models}

Let the network be represented by a directed graph $G = (V, E)$,
where $V$ is the set of nodes and $E$ is the set of links. For
each link $e \in E$, $c_e > 0$ is its capacity. A {\em multicast
session} consists of the source node and all the receivers
corresponding to a file. Let $s$ denote a session or the source of
a session interchangeably. In a session $s$, the data traffic is
routed along multiple multicast trees, each rooted at the source
$s$ and covering all the receivers. A multicast tree is a {\em
Steiner} tree; it may contain nodes not in the session, which are
called Steiner nodes. Let the set of all allowed multicast trees
for session $s$ be denoted by $T_s$. Throughout the paper, we
assume $T_s$ contains all possible multicast trees unless
specified otherwise. Let $S$ be the set of all multicast sessions.
Then, $T = \cup_{s \in S} T_s$ is the collection of all multicast
trees for all sessions. The multicast trees can be indexed in an
arbitrary order as $t_1, t_2, \cdots, t_{|T|}$, where $|\cdot|$ is
the cardinality of a set. Though $|T|$ is finite, it might be
exponential in the number of links. Let $x_s$ be the flow rate of
session $s \in S$ and $y_t$ be the flow rate of a multicast tree
$t$. We have $x_s = \sum_{t \in T_s} y_t$.
Finally, let $x$ and $y$ denote the vectors of $x_s$ and $y_t$,
respectively.

Each session $s \in S$ is associated with a utility function
$U_s(x_s), 0 \leq m_s \leq x_s \leq M_s$.
The assumption on the utility functions is, for every $s \in S$,
\beit
\item $A1$:
$U_s$ is well-defined (real-valued), non-decreasing, strictly concave
on $[m_s, M_s]$, and twice continuously differentiable on
$(m_s, M_s)$.
\eeit
Assumption $A1$ is widely made in the literature (see \cite{LL99}
\cite{KMT98}). The family of such functions is flexible enough for achieving a wide variety of network objectives, e.g., proportional fairness.
It is further justified by the fact that the most pervasive rate-control protocols in use, i.e., various versions of TCP, are shown
to optimize different strictly concave objective functions \cite{Low03}.


The problem is to find the optimal resource (i.e., session and
multicast-tree rates) allocation to maximize the sum of session utilities
under the capacity constraints and session rate constraints.
We call the optimization problem the master problem (MP), which is as follows.
\bea
\label{formula:primal}
& \max \ f(x, y) = \sum_{s \in S} U_s(x_s) & \\
\mbox{s.t.} & x_s = \sum_{t \in T_s} y_t, & \forall s \in S \nonumber \\
& \sum_{t \in T: e \in t} y_t \leq c_e, & \forall e \in E
\label{ineq:capacity_constr} \\
& m_s \leq x_s \leq M_s, & \forall s \in S \nonumber \\
& y_t \geq 0, & \forall t \in T. \nonumber
\eea
In the optimization problem (\ref{formula:primal}),
the session flow rates $x$ and the tree flow rates $y$
are the decision variables.
Without loss of generality, we make a technical assumption on the problem
(\ref{formula:primal}).
\beit
\item $A2$: There exists a feasible solution $(\bar{x}, \bar{y})$
such that $m_s \leq \bar{x}_s \leq M_s$ for any session $s \in S$
and (\ref{ineq:capacity_constr}) holds with strict inequality at
$(\bar{x}, \bar{y})$.
\eeit

Let $\lambda_e$ be the Lagrangian multiplier associated with
the constraint (\ref{ineq:capacity_constr}). The
Lagrangian function is
\begin{align}
\label{eq:lagrangian}
L(x, y, \lambda)
= & \sum_{s \in S} U_s(x_s)
+ \sum_{e \in E} \lambda_e (c_e - \sum_{t \in T: e \in t} y_t) \nonumber \\
= & \sum_{s \in S} (U_s(x_s) - \sum_{t \in T_s} y_t \sum_{e \in t} \lambda_e)
+ \sum_{e \in E} \lambda_e c_e.
\end{align}
Note that the Lagrangian function
$L(x, y, \lambda)$ is strictly concave in $x$, but linear
in $y$.

The dual function is
\bea
\label{eq:dual_fun}
\theta(\lambda)  \triangleq &
\max \ \ \ L(x, y, \lambda) & \\
\mbox{s.t.} & x_s = \sum_{t \in T_s} y_t, & \forall s \in S \nonumber \\
& m_s \leq x_s \leq M_s, & \forall s \in S \nonumber \\
& y_t \geq 0, & \forall t \in T. \nonumber
\eea
In the Lagrangian maximization
problem, the decision variables are
the vectors $x$ and $y$.

Now the dual problem of (\ref{formula:primal}) is
\bea
\label{formula:dual}
\mbox{Dual:} & \min & \theta (\lambda) \\
\mbox{s.t.} & \lambda \geq 0. \nonumber
\eea
In the dual problem (\ref{formula:dual}), the decision variables
are the vector $\lambda$.


\section{A Distributed Algorithm}
\label{sec:subgradient}

In this section, we will illustrate how the problem
(\ref{formula:primal}) can be solved by a distributed subgradient
algorithm.

\subsection{Subgradient Algorithm}
The dual problem (\ref{formula:dual}) can be solved by
a standard subgradient method
as in Algorithm \ref{algo:subgradient},
where $\delta_e(k)$ is a positive scalar step size, $[\cdot]_+$
and $[\cdot]_{m_s}^{M_s}$ denote the projection onto
the non-negative real numbers and the interval of $[m_s, M_s]$, respectively
 \cite{Bert99} \cite{BazaraaSS}.
There are two possible step size rules:
\beit
\item Rule I (Constant step size): $\delta_e(k) = \delta > 0$,
    for all time $k \geq K$ for some finite $K$.
\item Rule II (Diminishing step size): $\delta_e(k) \leq \delta_e(k-1)$
    for all time $k \geq K$ for some finite $K$.
    $\lim_{k \rightarrow \infty} \delta_e(k) = 0$
    and $\lim_{k \rightarrow \infty} \sum_{u = 0}^k \delta_e(u) = \infty$. Without loss
    of generality, we assume $\delta_e(k) \neq 0$ for any $k$.
\eeit

At the update (\ref{eq:sub_x}) and (\ref{eq:sub_y})
of Algorithm \ref{algo:subgradient}, we need to compute
a minimum-cost Steiner tree problem. Under any fixed
dual cost vector $\lambda \geq 0$, for any session $s \in S$, define a min-cost
Steiner tree by
\be
\label{eq:min_cost_tree}
t(s, \lambda) = \mbox{argmin}_{t \in T_s} \{\sum_{e \in t} \lambda_e \},
\ee
where the tie is broken arbitrarily. Because (\ref{eq:min_cost_tree})
is an optimization problem over all allowed trees, we call
(\ref{eq:min_cost_tree}) a {\em global min-cost tree problem}, and
the achieved minimum cost the {\em global minimum tree cost}.
We denote this global minimum tree cost under a fixed $\lambda \geq 0$ by
\be
\label{eq:min_cost_tree_cost}
\gamma(s, \lambda) = \sum_{e \in t(s, \lambda)} \lambda_e.
\ee

\begin{algorithm}
\caption{Subgradient Algorithm}
\label{algo:subgradient}
\begin{align}
& \lambda_e(k + 1)  = [\lambda_e(k) - \delta_e(k) (c_e
    - \sum_{t \in T: e \in t} y_t(k))]_+,
\forall e \in E \label{eq:sub_lambda} \\
& x_s(k + 1) = [(U_s')^{-1} (\gamma(s, \lambda(k + 1)))]_{m_s}^{M_s},
\forall s \in S  \label{eq:sub_x} \\
& y_t(k + 1) =
\left\{ \begin{array}{ll}
         x_s(k + 1) & \mbox{if $t = t(s, \lambda(k+1))$;}\\
         0 & \mbox{otherwise}, \end{array} \right.
\forall t \in T.  \label{eq:sub_y}
\end{align}
\end{algorithm}

\noindent {\bf Remark}: Algorithm \ref{algo:subgradient} is a
distributed algorithm. In order to compute the tree cost, each
link $e$ can independently compute its dual cost $\lambda_e$ based
on the local aggregate rate passing through the link. Then, the
tree cost can be accumulated by the source $s$ based on the link
cost values along the tree. To find the minimum-cost tree $t(s,
\lambda(k))$, each source needs to compute the minimum-cost
directed Steiner tree, which is an NP-hard problem. We will
address this issue in Section \ref{sec:colgen}. Other than that,
the subgradient algorithm is completely decentralized.

\subsection{Convergence Results}
Let $\Lambda^* = \{\lambda \geq 0:
\theta(\lambda) = \min_{\lambda \geq 0} \theta(\lambda) \}$
be the set of optimal dual variables.
Let $f^*$ be the optimal function value of the problem
(\ref{formula:primal}) and $(x^*, y^*, \lambda^*)$ denote one of
the optimal primal-dual solutions. Obviously, $f^*$ is bounded.

\begin{lemma}
\label{lemma:sub}
Under assumptions $A1$ and $A2$,
\beit
\item $(a)$ There is no duality gap between the primal problem (\ref{formula:primal})
and the dual problem (\ref{formula:dual}), i.e.,
$f^* = \theta(\lambda^*)$ for any $\lambda^* \in \Lambda^*$.
\item $(b)$ For any $\lambda \geq 0$, $(x, y)$ obtained by
(\ref{eq:sub_x}) and (\ref{eq:sub_y}) are one of the Lagrangian maximizers
with the Lagrangian multiplier $\lambda$. Furthermore,
$x$ obtained by (\ref{eq:sub_x}) is the unique Lagrangian maximizer.
\item $(c)$ For any $\lambda^* \in \Lambda^*$, the solution obtained by
(\ref{eq:sub_x}) is the unique optimal solution $x^*$ of (\ref{formula:primal}).
\item $(d)$ $\Lambda^*$ is a non-empty compact set.
\eeit
\end{lemma}
\IEEEproof{
See Appendix A.
}


\begin{theorem}
\label{theorem:convergence_sub_lambda_x}
Let $d(\lambda, \Lambda^*) = \min_{\lambda^* \in \Lambda^*} ||\lambda - \lambda^*||$.
For any $\epsilon > 0$, there exists some $\delta_0 > 0$ such that
\beit
\item
if the sequence of step size $\{ \delta(k) \}$ satisfies
step size rule I and $\delta(k) < \delta_0$ for all $k$,
\item
or if the sequence of step size $\{ \delta(k) \}$
satisfies step size rule II,
\eeit
then there exists a sufficiently
large $K_0 < \infty$ such that,
with any initial $\lambda(0) \geq 0$, we have
$d(\lambda(k), \Lambda^*) < \epsilon$ and $||x(k) - x^*|| < \epsilon$ for all $k \geq K_0$.
\end{theorem}
\IEEEproof{
See Appendix A. (See also \cite{LS06A} for a similar situation.)
}


We now discuss the convergence of the tree rate vector $y(k)$. The
difficulty of proving the convergence of $y(k)$ arises from the
linearity of the Lagrangian function in (\ref{eq:lagrangian}) on
the vector $y$, and there is no standard result about the
convergence of $y(k)$. In fact, the tree rate vector $y(k)$ does
not converge in the normal sense \cite{Wang03canshortest-path}.
From the update (\ref{eq:sub_y}), we see that each source only
uses one tree (i.e., assigns a positive rate) each time and shifts
flow from one tree to another from time to time. We further
noticed that, by pushing the session traffic onto only one tree at
a time, the link capacity constraints are often violated. This
means that the rate allocation on each time slot may not even be
feasible.
In the following, we will show that the tree rates
converge to the optimal values in the time average sense and that
the time-average link flow rates satisfy the link capacity constraints.


\begin{theorem}
\label{theorem:convergence_sub_y}
For any link $e$ and a finite time $k_0$, there exists a constant $M_e < \infty$\footnote{$M_e$ only depends on $k_0$ and is independent of $k$.}
such that for any time $k$,
\be
\sum_{u = k_0}^k \sum_{t \in T: e \in t} y_t(u)
\leq c_e (k - k_0 + 1)
+ M_e. \nonumber
\ee
\end{theorem}
\IEEEproof{
See Appendix A.
}

Let $H$ denote the $|E| \times |T|$ link-tree incidence matrix where
$[H]_{et} = 1$ if $e \in t$; otherwise, $[H]_{et} = 0$.
Let $A$ denote the $|S| \times |T|$ session-tree incidence matrix where $[A]_{st} = 1$ if $t \in T_s$;
otherwise, $[A]_{st} = 0$.
For an arbitrary $k_0$,
let us define a sequence $\{\bar{y}(k)\}_{k \geq k_0}$, where
\be
\label{eq:def_y_bar}
\bar{y}(k) = \frac{\sum_{u = k_0}^k y(u)}{k - k_0 + 1}.
\ee

\begin{theorem}
\label{theorem:avg_y_feasible}
\be
\lim_{k \rightarrow \infty} \sup H \bar{y}(k) \leq c.
\nonumber
\ee
Furthermore, for any limit point $\bar{y}^*$ of the
sequence $\{ \bar{y} (k) \}$,
$H \bar{y}^* \leq c$.
\end{theorem}
\IEEEproof{
See Appendix A.
}

For any $\epsilon > 0$, let us define
$\mathcal{Y}^* (\epsilon) = \{y \geq 0: H y \leq c,
||A y - x^*|| \leq \epsilon \}$.
When $\epsilon = 0$,
$\mathcal{Y}^* (0) = \mathcal{Y}^*
= \{y \geq 0: H y \leq c,
A y = x^* \}$, which is the set of optimal tree rate allocation.
\begin{theorem}
\label{theorem:convergence_avg_y}
For any $\epsilon > 0$,
with any initial $\lambda(0) \geq 0$,
every limit point of the sequence $\{ \bar{y} (k) \}$ is in the set
$\mathcal{Y}^* (\epsilon)$.
\end{theorem}
\IEEEproof{
See Appendix A.
}

\noindent {\bf Remark 1}:
By Theorem \ref{theorem:convergence_avg_y}, the time average of the tree
rate vectors, $\bar{y}(k)$, converges to the optimal set.
Theorem \ref{theorem:convergence_sub_y} to Theorem
\ref{theorem:convergence_avg_y} hold under both the step size rule
I and II.

\noindent {\bf Remark 2}:
Algorithm \ref{algo:subgradient} can be viewed as a congestion control algorithm.
On each time slot, the algorithm looks for a multicast tree whose overall congestion price (which is the sum of the link costs on the tree) is small and uses such a tree to deliver data. If a link is very congested, its link cost (queue size) is very high and it is less likely to be selected as part of a multicast tree.
Hence, our algorithm automatically adapts to network congestion and avoids the buildup of very large queues. The buffer size requirement for preventing buffer overflow will likely to be moderate. There are also variants of the subgradient algorithm which update
the instantaneous flow rates gradually, avoiding sending big bursts of data to the link queues. The queue sizes can be reduced greatly with such variants.




\section{Column Generation Method With Imperfect
Global Min-cost Tree Scheduling}
\label{sec:colgen}

In Section \ref{sec:subgradient}, we have described a distributed
algorithm to solve the master problem (\ref{formula:primal}). However, it is not practical enough since the subproblem (\ref{eq:min_cost_tree}) is NP-hard \cite{HRW92}.
The column generation method can be introduced to reduce the number of
times that the min-cost Steiner tree subproblem is invoked. We
also consider applying imperfect tree scheduling, which are
approximate or heuristic, sub-optimal solutions to the Steiner
tree subproblem.
The column generation method with approximation
was also proposed in \cite{ZhengCXF09} to solve the problem of
wireless link scheduling.
In \cite{ZhengCXF09}, the optimization problem
has a similar constraint as in (2),
which is that the aggregate link flow rate is no more than
the link capacity. Since only singe-path routing is considered in \cite{ZhengCXF09}, the aggregate link flow rate
can be expressed explicitly as the sum of the path flow rates.
However, to express the link flow rate
explicitly in this paper requires enumerating all
multicast trees, the number of which increases dramatically as the network
size increases. On the other side of the constraint inequality, the
link capacity is a known constant in the current problem. In \cite{ZhengCXF09},
the link capacities can be expressed as a convex combination of all link schedules where the coefficients are decision variables; this expression
requires enumerating all possible wireless schedules, the number of which can be enormously large.
The distinctions make the two problems
sufficiently different and
all the results in this paper need to be derived independently.

\subsection{Column Generation Method}

The main idea of column generation is to start with a
subset of the tree set $T$
and bring in new trees only when needed. Consider a subset of
$T$ containing only a small number of trees, i.e.,
$T^{(q)} = \{t_i \in T: \forall i = 1,
\cdots, q \}$.
We make sure that $T^{(q)}$ contains at least one
tree for each source $s$.
Denote $T_s^{(q)}$ the subset of trees in $T^{(q)}$ that are
rooted at source $s$, i.e.,
$T_s^{(q)} = \{t: t \in T^{(q)} \cap T_s \}$.
We can formulate the following
restricted master problem (RMP) for $T^{(q)}$,
which will be called the $q^{th}$-RMP.
\bea
\label{formula:primal_RMP}
\mbox{$q^{th}$-RMP: } & \max \ \sum_{s \in S} U_s(x_s) & \\
\mbox{s.t.} & x_s = \sum_{t \in T_s^{(q)}} y_t,
& \forall s \in S \nonumber \\
& \sum_{t \in T^{(q)}: e \in t} y_t \leq c_e,
& \forall e \in E \label{ineq:capacity_constr_RMP} \\
& m_s \leq x_s \leq M_s, & \forall s \in S \nonumber \\
& y_t \geq 0, & \forall t \in T^{(q)}. \nonumber
\eea
The value of $q$ is usually small and the trees in the set
$T^{(q)}$ can be examined one-by-one.
The Lagrangian function, the dual function, and the dual problem
of the $q^{th}$-RMP can be formulated similarly as in
(\ref{eq:lagrangian}), (\ref{eq:dual_fun}), and (\ref{formula:dual}),
where the set $T$ is replaced by the set $T^{(q)}$.


The $q^{th}$-RMP is more restricted than the MP. Thus, any optimal
solution to the $q^{th}$-RMP is feasible to the MP and provides a
lower bound of the optimal value of the MP. By gradually
introducing more trees (columns) into $T^{(q)}$ and expanding the
subset $T^{(q)}$, we will improve the lower bound of the MP
\cite{BVY03, JX06, KWE07}.

\subsection{Apply the Subgradient Algorithm to the RMP}

The distributed subgradient algorithm can
be used to solve the $q^{th}$-RMP.
Here, we define the following problem of finding the min-cost tree
$t^{(q)}(s, \lambda)$ under the
link cost vector $\lambda \geq 0$.
\be
\label{eq:min_cost_tree_local}
t^{(q)}(s, \lambda) = \mbox{argmin}_{t \in T_s^{(q)}} \{\sum_{e \in t} \lambda_e \},
\ee
The optimization is taken over the $|T_s^{(q)}|$ currently known trees.
The problem in (\ref{eq:min_cost_tree_local}) is called the
{\em local min-cost tree problem}, and the achieved minimum
cost is called the {\em local minimum tree cost}.
We denote this local minimum cost under $\lambda \geq 0$
by
\be
\label{eq:min_cost_tree_cost_local}
\gamma^{(q)}(s, \lambda) = \sum_{e \in t^{(q)}(s, \lambda)} \lambda_e.
\ee
If there is more than one tree achieving the local minimum cost,
the tie is broken arbitrarily.

\subsection{Introduce One More Tree (Column)}

Now the question is how to check whether the optimum of the
$q^{th}$-RMP is optimal for the MP, and if not, how to introduce a
new column (tree). It turns out there is an
easy way to do both. Let $(\bar{x}^{(q)}, \bar{y}^{(q)}, \bar{\lambda}^{(q)})$
denote one of the optimal primal-dual solutions of the
$q^{th}$-RMP.

\begin{lemma}
\label{fact:RMP_opt_cond}
$(\bar{x}^{(q)}, \bar{y}^{(q)}, \bar{\lambda}^{(q)})$ is
optimal to the MP if and only if
$h_s(\gamma(s, \bar{\lambda}^{(q)})) = h_s(\gamma^{(q)} (s, \bar{\lambda}^{(q)}))$,
for all $s \in S$, where
\be
h_s(w) = U_s([(U'_s)^{-1}(w)]_{m_s}^{M_s})
    - [(U'_s)^{-1}(w)]_{m_s}^{M_s} \cdot w, \ w \geq 0. \nonumber
\ee
\end{lemma}
\IEEEproof{
See Appendix B.
}

From Lemma \ref{fact:RMP_opt_cond}, a sufficient condition for optimality is that the local minimum tree cost
is equal to the global minimum tree cost,
i.e., $\gamma(s, \bar{\lambda}^{(q)})= \gamma^{(q)} (s, \bar{\lambda}^{(q)})$.
We state the rule of introducing a new column in the following.
\begin{fact}
\label{fact:RMP_intro_col_rule}
Any tree achieving a cost less than the local
minimum tree cost could enter the subset $T^{(q)}$
in the RMP.
The tree achieving the
global minimum tree cost is one possible candidate and is
often preferred \cite{ZhengCXF09}.
\end{fact}


\subsection{Column Generation by Imperfect Global Tree scheduling}

The min-cost Steiner tree subproblem (\ref{eq:min_cost_tree}) is
NP-hard, which makes the step of column
generation very difficult. We now consider approximation algorithms to this
subproblem. We may solve it approximately, and this is
referred as {\em imperfect global tree scheduling}.\footnote{Note that the local min-cost tree problem
(\ref{eq:min_cost_tree_local}) can be easily solved precisely
since the number of extreme points (i.e., candidate trees) of $T^{(q)}$ is
usually small, and hence, enumerable.}

Suppose we are able to solve (\ref{eq:min_cost_tree})
with an approximation ratio $\rho \geq 1$, i.e.,
\be
\label{ineq:tree_approx_ratio}
\frac{1}{\rho} \gamma_{\rho} (s, \lambda) \leq
\gamma (s, \lambda),
\ee
where $\gamma_{\rho} (s, \lambda)$ is the
cost of the tree given by the approximate solution.

\subsubsection{A $\rho$-Approximation Approach}

We develop a column generation method with imperfect global
min-cost tree scheduling as follows. Later, we will show a
guaranteed performance bound of this approach. Algorithm
\ref{algo:colgen_imperfect} in fact describes a whole class of
algorithms, depending on the parameter $\Delta$ and $\rho$,
representing different performance, convergence speed and complex
tradeoffs. More detailed comments about the property of this class
of algorithms can be found in \cite{ZhengCXF09}.
\begin{algorithm}
\caption{Column Generation with Imperfect Global Tree Scheduling}
\label{algo:colgen_imperfect}
\beit
\item Initialize: Start with a collection of $T^{(q)}$ trees. (Assume
    Assumption $A2$ holds for the $q^{th}$-RMP.)
\item Step $1$:
        Run the subgradient algorithm
        (\ref{eq:sub_lambda})-(\ref{eq:sub_y}) $\Delta$
        (typically several) times on the $q^{th}$-RMP.
\item Step $2$: For each source $s$, solve the global min-cost tree problem
        (\ref{eq:min_cost_tree}) {\em with an
        approximation ratio $\rho$} under
        the current dual cost $\lambda$.
        \beit
        \item
        If the tree corresponding to
        the {\em approximate solution} of (\ref{eq:min_cost_tree})
        is already in the current collection of trees, do nothing;
        \item Otherwise,
        introduce this tree
        into the current collection of trees, and
        increase $q$ by $1$.
        \eeit
        Go to Step $1$.
\eeit
\end{algorithm}

\noindent {\bf Remark 1}: In step 1, note that we do not have to
run the subgradient algorithm on the $q^{th}$-RMP until
convergence. The number $\Delta$ can be as small as 1, in which
case the algorithm degenerates into the pure subgradient
algorithm, Algorithm \ref{algo:subgradient}, but with
approximation in the tree computation. In this case, the global
min-cost tree problem is solved in every iteration. On the other
hand, if $\Delta$ is large, then the subgradient algorithm on the
$q^{th}$-RMP runs until near convergence. This will be a pure
column generation algorithm. In this case, the global min-cost
tree problem is solved after many iterations. However, the entire
algorithm may take more iterations to complete. Hence, $\Delta$
determines how often the global min-cost tree is computed. It can
be chosen to trade-off the complexity of solving the global
min-cost tree problem and the number of iterations.

\noindent {\bf Remark 2}: If the approximate tree derived in
step 2 has a higher tree cost than that of the local min-cost tree
among the existing trees already selected, we define the existing
tree with the lowest cost as the solution to the approximation
algorithm. Hence, the cost of the imperfect (approximate) tree
cannot be higher than any of the existing trees.

\noindent {\bf Remark 3}:
Note that once a tree enters the set $T^{(q)}$,
it will not be removed from the set.
Since the set $T^{(q)}$
is usually small, it is possible for a source to
manage its current collection of trees.
Furthermore, if the global min-cost tree
problem (\ref{eq:min_cost_tree}) can be solved approximately in a
decentralized fashion, then Algorithm \ref{algo:colgen_imperfect}
is completely decentralized. In Section \ref{section:experiment},
we will introduce some approximation algorithms for the min-cost
Steiner tree problem.


\subsubsection{Convergence with Imperfect Global Tree Scheduling}
\begin{theorem}
\label{theo:convergence_colgen}
There exists a $q$, $1 \leq q \leq |T|$, such that
Algorithm \ref{algo:colgen_imperfect}
converges to one optimal primal-dual solution
of this particular $q^{th}$-RMP, i.e.,
$(\bar{x}^{(q)}, \bar{\lambda}^{(q)})$.
Furthermore, after
Algorithm \ref{algo:colgen_imperfect} converges to
$(\bar{x}^{(q)}, \bar{\lambda}^{(q)})$,
$\gamma_{\rho} (s, \bar{\lambda}^{(q)}) = \gamma^{(q)} (s, \bar{\lambda}^{(q)})$
for any source $s \in S$.
\end{theorem}
\IEEEproof{
See Appendix B.
}

\subsubsection{Performance Bound under Imperfect Tree Scheduling}

Theorem \ref{theo:convergence_colgen} says that the column
generation method with imperfect global tree scheduling
converges to a sub-optimum of the MP. Next, we will
prove that the performance of this sub-optimum is bounded.
We make assumptions $A3$ and $A4$.

\beit
\item $A3$: For any source $s \in S$, $m_s \geq 0$ is sufficiently
small such that, if the column generation method
with imperfect global tree scheduling converges to
$(\bar{x}^{(q)}, \bar{\lambda}^{(q)})$
on the $q^{th}$-RMP, then $\bar{x}_s^{(q)} > m_s$.
\item $A4$: $U_s(m_s) - m_s \cdot U'_s(m_s) \geq 0, \forall s \in S$.
\eeit

\noindent {\bf Remark:} Assumption $A4$ is not very restricting. For instance, it
will hold if $U_s \geq 0$, concave, non-decreasing and
differentiable on $[0, M_s]$. The latter three conditions are already implied by Assumption $A1$
but on $[m_s, M_s]$ only.

\begin{theorem} [Bound of Imperfect Global Tree Scheduling]
\label{theo:colgen_bound}
Under the additional assumptions $A3$ and $A4$,
if the column generation method
with imperfect global tree scheduling converges to
$(\bar{x}^{(q)}, \bar{\lambda}^{(q)})$
on the $q^{th}$-RMP, we have
\be
\label{ineq:approx_dual_obj_bound}
\theta^{(q)} (\bar{\lambda}^{(q)})
\leq \sum_{s \in S} U_s(x_s^*)
\leq \theta (\rho \bar{\lambda}^{(q)})
\leq \rho \theta^{(q)} (\bar{\lambda}^{(q)}).
\ee
\end{theorem}
\IEEEproof{
See Appendix B.
}

Since the strong duality holds on the $q^{th}$-RMP,
$\sum_{s \in S} U_s(\bar{x}_s^{(q)}) = \theta^{(q)} (\bar{\lambda}^{(q)})$,
we have the following.
\begin{corollary} [$\rho$-Approximation Solution to the MP]
\label{corollary:approx_bounds}
Under the additional assumptions $A3$ and $A4$, we have
\be
\label{ineq:approx_obj_bound}
\sum_{s \in S} U_s(\bar{x}_s^{(q)})
\leq \sum_{s \in S} U_s(x_s^*)
\leq \rho \sum_{s \in S} U_s(\bar{x}_s^{(q)}).
\ee
If $\rho = 1.0$,
(\ref{ineq:approx_obj_bound}) holds with equality,
then Algorithm \ref{algo:colgen_imperfect}
is the column generation method with perfect global min-cost tree scheduling,
and it converges to one optimum of MP.
\end{corollary}

Corollary \ref{corollary:approx_bounds} says that
the column generation method with imperfect global
tree scheduling converges to a sub-optimum of the MP and
achieves an approximation ratio no less than the reciprocal of the approximation ratio
to the global min-cost tree problem.


\noindent {\bf Remark:} Possible utility functions include
$U_s(x_s) = w_s \ln(x_s + e)$ and
$U_s(x_s) = \frac{w_s}{1-\beta} x_s^{1-\beta}$,
where $0 < \beta < 1$ and $w_s > 0$.




\section{Illustrative examples}
\label{section:experiment}

In this section, we give illustrative examples showing the effect
of universal swarming and the performance of our algorithms.



\subsection{Universal Swarming versus Separate Swarming}

\begin{figure}[t]
\begin{center}
\includegraphics[width=3.4in]{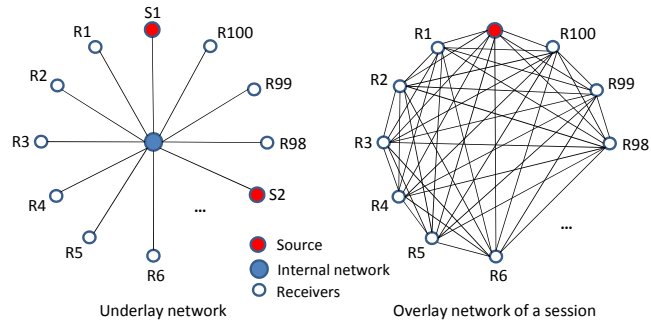}
\end{center}
\caption{Network topology used in the simulation.}
\label{fig:topology}
\end{figure}

We test our algorithms in various scenarios by varying the sizes
of the resource-rich and resource-poor sessions and the locations
of bandwidth bottleneck. We have nine test cases (profiles) where
we assume the internal network has large enough capacity so that it
cannot be the bottleneck; the bottleneck lies on the
access links. Fig. \ref{fig:topology} shows the network topology used in the simulation. The underlay network topology is conceptually equivalent to a star, where the internal network is represented as a node in the center. The overlay network of a session is a complete graph where every pair of nodes has an incoming and an outgoing link between them. Our algorithm operates on the overlay network.
In each of the profiles $A1$, $A2$ and $A3$, there
is a large resource-rich session (RRS) and a small resource-poor
session (RPS); in each of the profiles $B1$, $B2$ and $B3$, there
is an RRS and an equal-sized RPS; and in each of the profiles
$C1$, $C2$ and $C3$, there is a small RRS and a large RPS. Each
large session contains $90$ receivers; each small session contains
$10$ receivers; and each medium session contains $50$ receivers.
Each session has a single source. We also vary the bottleneck location
of the RRS session so that we can examine how intersession cooperation
affects the rate allocation in each case. In profiles $A1$, $B1$
and $C1$, the bottleneck of the RRS is at the download links; in
profile $A2$, $B2$ and $C2$, the bottleneck of the RRS is at the
upload link of its source; and in profile $A3$, $B3$ and $C3$, the
RRS is bottlenecked by its aggregate upload bandwidth. In all
cases, the RPS is bottlenecked at its aggregate upload bandwidth.
Note that if the bottleneck of the RPS is at its source upload
link or the receiver download links, then there is no way to
improve its session rate.

In the simulation, we use $U_s(x_s) =
\ln(x_s + e)$ as the utility function, and run the subgradient
algorithm for $10000$ iterations so that we reach convergence for
all the cases.
In most cases, after $2000$ iterations,
the algorithm has already produced a
rate allocation very close to the final result.
Furthermore, the distribution considered in this paper takes place in a managed, relatively static network environment with constant link bandwidth, long-lasting sessions and no other background traffic. We do not have the usual issues
associated with end-system P2P file-sharing environment, such
as high churn rates and dynamic background traffic.
Hence, the number of iterations required to reach
convergence is affordable.
The step size rules and the initial step sizes used
in profiles are slightly different from each other.
In our simulation, we have selected proper step sizes for the test cases by a few trials so that convergence occurs within $10000$ iterations. It is hard to apply
the same step size rule for all the profiles and reach convergence within $10000$ iterations. For test cases $A1$-$A3$ and $B1$-$B3$, we use the diminishing step size rule with $\delta_e(0) = \delta$ and $\delta_e(k) = \delta_e(k-1)/ \sqrt{k}$ for $k \geq 1$. The initial step size $\delta$ varies between $5 \times 10^{-8}$ and $5 \times 10^{-7}$. For the other test cases, we use the constant step size rule with $\delta_e(k)= 5 \times 10^{-7}$.

In each test case, we compare the rate allocation results of
separate swarming with that of universal swarming. For the
separate swarming, we use a minimum spanning tree algorithm for
the subproblem to compute the global min-cost tree. This is
possible since the sessions are separated from each other and the
overlay network for each session contains no Steiner nodes.
On the other hand, for universal swarming,
we use the algorithm by Charikar {\em et. al}
with tree level $2$, as proposed in
\cite{Charikar98}, to find approximate
minimum-cost trees
\footnote{
We implemented the algorithm by Charikar et. al in a centralized fashion. If we had a distributed implementation, our algorithm would have been fully distributed. However, to the best of our knowledge, there has not been any distributed algorithm for the min-cost Steiner tree problem that can provide bounded
performance on {\em any} directed graph. But, there exist distributed approximate
algorithms that have good performance in most cases but unbounded worst-case performance.
The distributed spanning tree algorithm is one such
example \cite{Humblet83}. Such a distributed Steiner tree algorithm may work well enough for the network graphs encountered in practice and can be incorporated into our overall rate-allocation algorithm.
}.
The algorithm achieves an approximation ratio
$i(i-1)|R_s|^{1/i}$ with time complexity $O(|V|^i|R_s|^{2i})$ for any level $i > 1$, where $|R_s|$ is
the number of receivers of session $s$ and $|V|$ is the number of nodes in the network.


\begin{table}[t]
\caption{Comparison of rate allocation between separate swarming and universal swarming}
\label{tab:experimentResult}
\begin{center}
\begin{tabular}{|c|c|c|c|c||c|c|}
\hline
\multicolumn{2}{|c|}{Test cases} & \multicolumn{3}{|c||}{Link bandwidth} & \multicolumn{2}{|c|}{Rate allocation} \\
\hline
Profile & Session & $u_s$ & $u_i$ & $d_i$
 & Separate & Universal \\
\hline
\multirow{2}{*}{A1}
    & Large RRS & 640   & 360   & 360   & 360   & 329.5 \\
    & Small RPS & 640   & 36    & 360   & 100   & 359.7 \\
\hline
\multirow{2}{*}{A2}
    & Large RRS & 280   & 360   & 360   & 280   & 280   \\
    & Small RPS & 280   & 36    & 360   & 64    & 280   \\
\hline
\multirow{2}{*}{A3}
    & Large RRS & 640   & 200   & 360   & 207   & 170.2 \\
    & Small RPS & 640   & 20    & 360   & 84    & 360   \\
\hline
\multirow{2}{*}{B1}
    & Medium RRS    & 640   & 360   & 360   & 360   & 192.5 \\
    & Medium RPS    & 640   & 36    & 360   & 48.8  & 188.7 \\
\hline
\multirow{2}{*}{B2}
    & Medium RRS    & 280   & 360   & 360   & 280   & 185.6 \\
    & Medium RPS    & 280   & 36    & 360   & 41.6  & 181.9 \\
\hline
\multirow{2}{*}{B3}
    & Medium RRS    & 640   & 200   & 360   & 212.8 & 112.7 \\
    & Medium RPS    & 640   & 20    & 360   & 32.8  & 110.4 \\
\hline
\multirow{2}{*}{C1}
    & Small RRS & 640   & 360   & 360   & 360   & 353   \\
    & Large RPS & 640   & 36    & 360   & 43.1  & 50.6  \\
\hline
\multirow{2}{*}{C2}
    & Small RRS & 280   & 360   & 360   & 280   & 276.8   \\
    & Large RPS & 280   & 36    & 360   & 39.1  & 50  \\
\hline
\multirow{2}{*}{C3}
    & Small RRS & 640   & 200   & 360   & 264   & 263.8 \\
    & Large RPS & 640   & 20    & 360   & 27.1  & 27.1  \\
\hline
\end{tabular}
\end{center}

\end{table}

Table \ref{tab:experimentResult} summarizes the simulation results
for our test cases. Let $u_s$, $u_i$, and $d_i$ be the source
upload bandwidth, and each receiver's upload and download
bandwidth, respectively. First, the simulation results show that
the subgradient algorithm always achieves the optimal rate
allocation in the separate swarming. Note that in the separate
swarming cases, we can easily check the optimality by simply
computing optimal rate allocation for each test case. The optimal
rate of each session can be computed as $\min\{ u_s, \min_{1 \leq
i \leq L} d_i, (u_s + \sum_{1 \leq i \leq L} u_i)/L \}$ where $L$
is the number of receivers\cite{KR06}. Second, the table shows
that with the universal swarming, the RPS can obtain the excess
resource of the RRS at small expense of the RRS. When the small
RPS is combined with the large RRS, its session rate improves
significantly while the large RRS loses a bit of its session rate.
When the session sizes of the RRS and RPS are the same, the
resulting session rates tend to be equalized,
which is partially due to the specific utility function used
in the simulation.
The assignment of the utility function $U_s(x_s) = \log(x_s + e)$ leads to proportional fairness between the sessions.
When the large RPS is combined with the small RRS, its
session rate still improves slightly with negligible impact on the
small RRS; this is also desirable since the RRS should not give up
its resource if it is not sufficiently abundant.

In summary, the results in Table \ref{tab:experimentResult}
indicate that universal swarming can indeed live up to the goal of
transferring excess resource from rich sessions to poor sessions
with very minor degradation to the rich sessions. One should keep
in mind that, in these test cases, universal swarming improves
over separate swarming even though we only compute sub-optimal
solutions for the universal swarming, whereas we compute optimal
solutions for the separate swarming. The potential improvement can
be even greater.

\subsection{Algorithm Performance}

\begin{figure}[t]
\begin{center}
\begin{minipage}{2in}
\begin{center}
\includegraphics[width=2in]{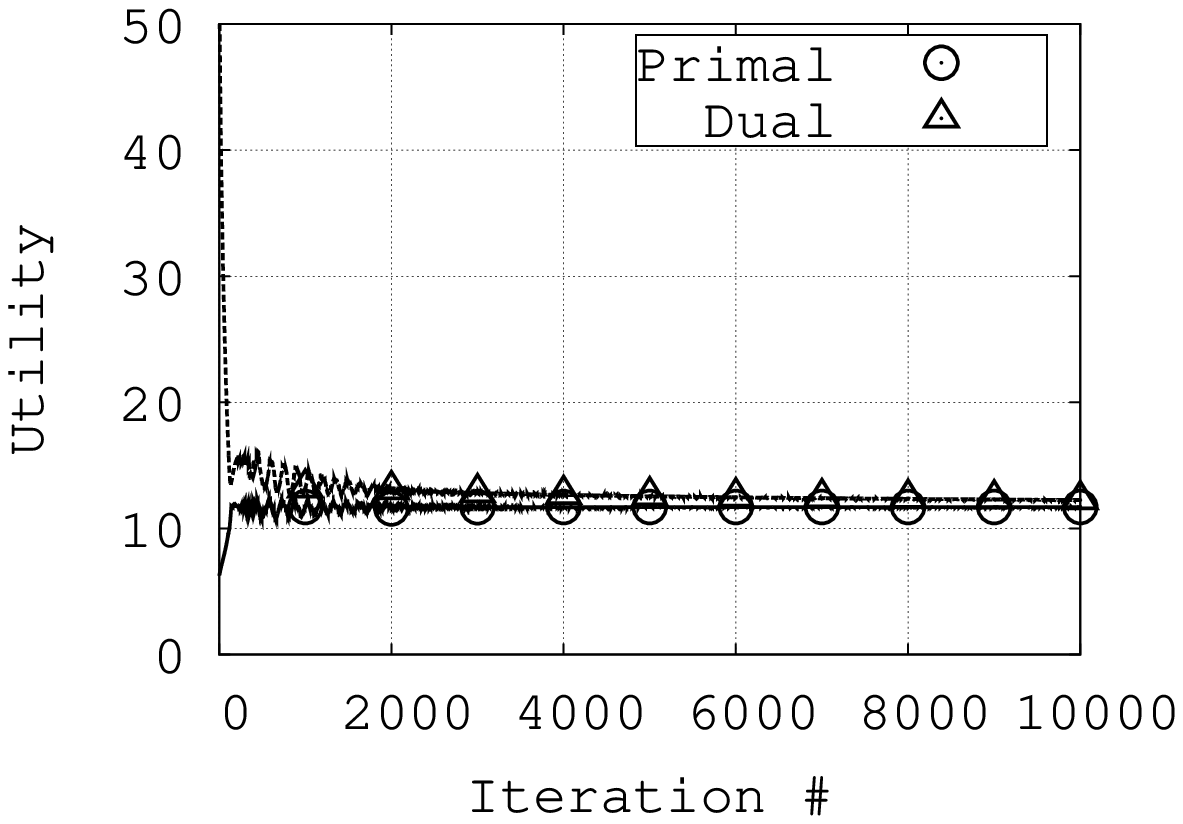} \\ (a) Objective values of A1
\label{fig:A1obj}
\end{center}
\end{minipage}
\begin{minipage}{2in}
\begin{center}
\includegraphics[width=2in]{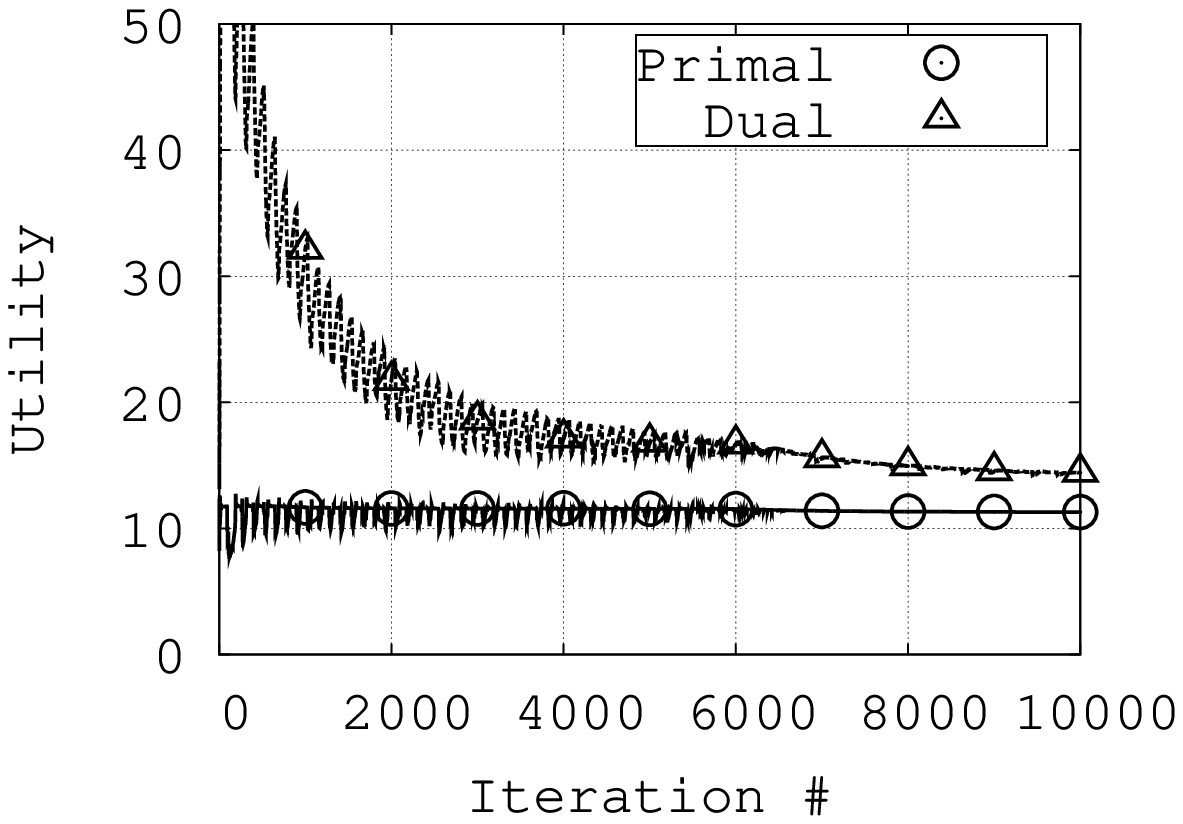} \\ (b) Objective values of A2
\label{fig:A2obj}
\end{center}
\end{minipage}
\begin{minipage}{2in}
\begin{center}
\includegraphics[width=2in]{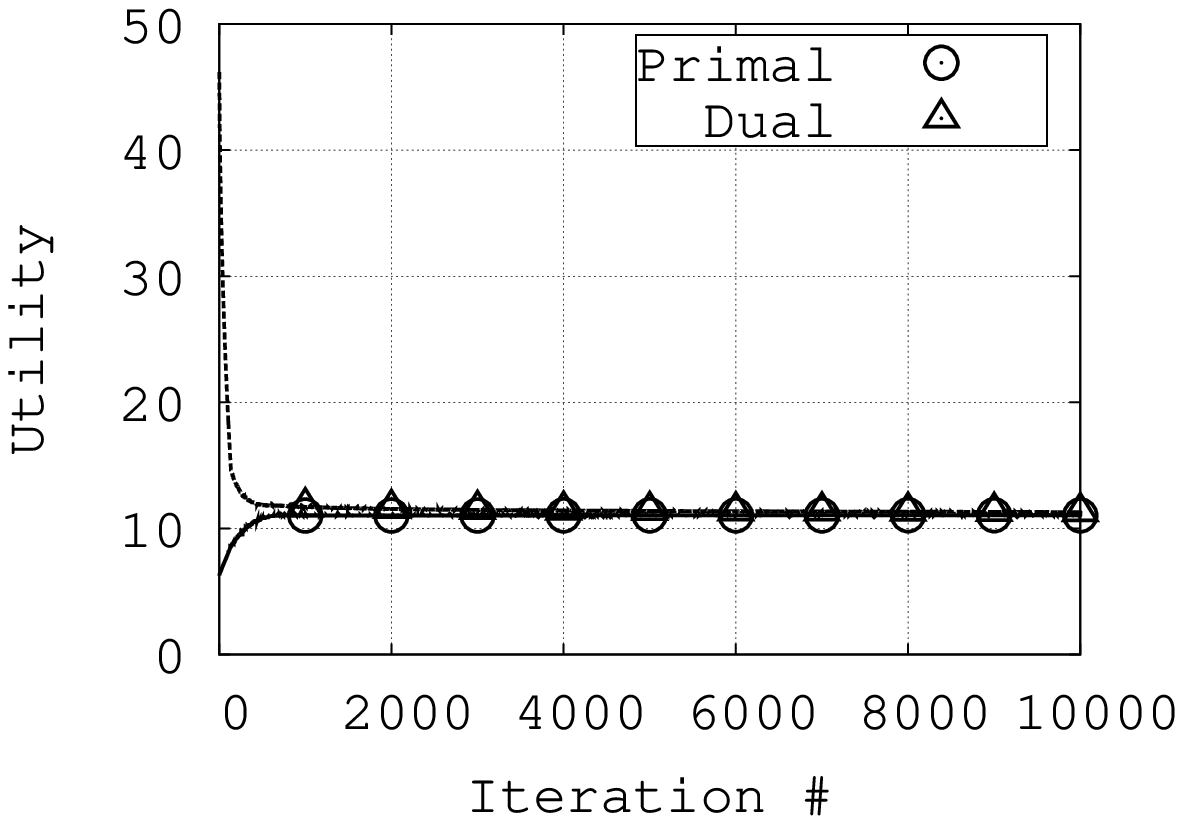} \\ (c) Objective values of A3
\label{fig:A3obj}
\end{center}
\end{minipage}
\end{center}
\begin{center}
\begin{minipage}{2in}
\begin{center}
\includegraphics[width=2in]{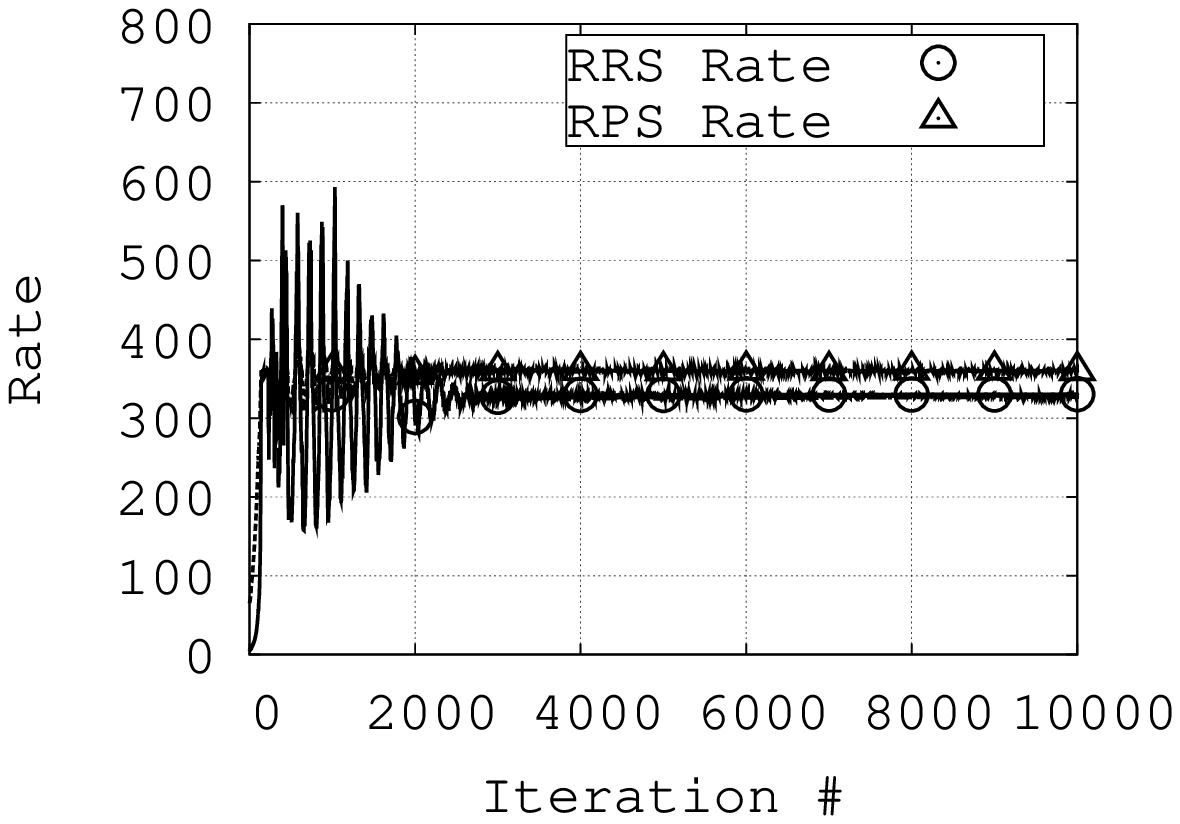} \\ (d) Rates of A1
\label{fig:A1rates}
\end{center}
\end{minipage}
\begin{minipage}{2in}
\begin{center}
\includegraphics[width=2in]{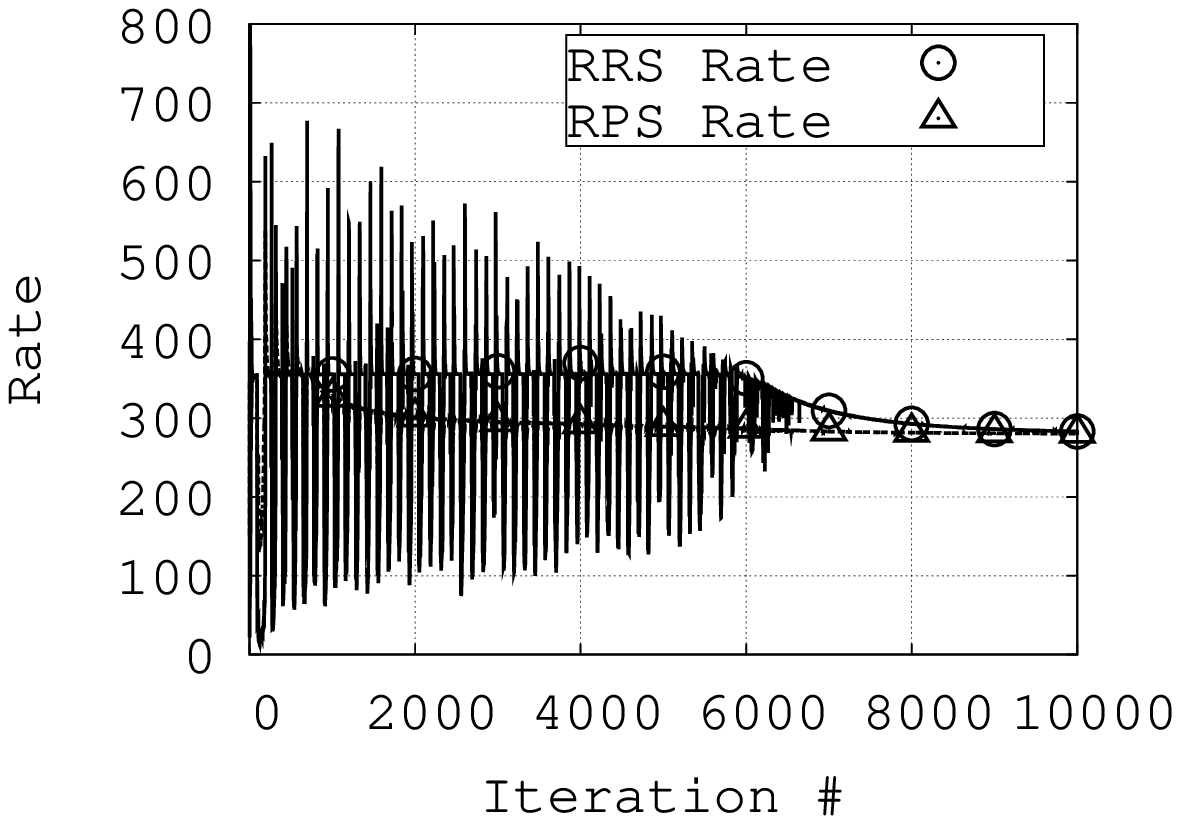} \\ (e) Rates of A2
\label{fig:A2rates}
\end{center}
\end{minipage}
\begin{minipage}{2in}
\begin{center}
\includegraphics[width=2in]{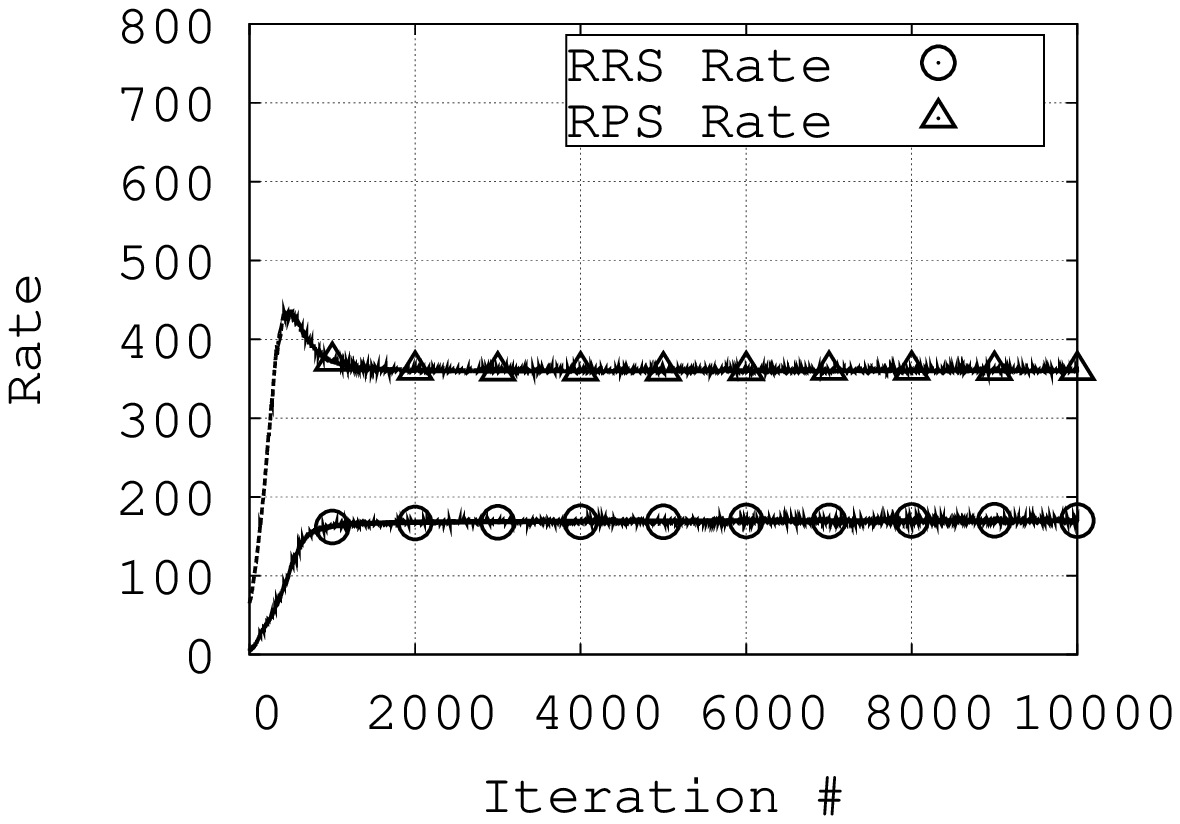} \\ (f) Rates of A3
\label{fig:A3rates}
\end{center}
\end{minipage}
\end{center}
\caption{Convergence of the pure subgradient algorithm, i.e., $\Delta = 1$.}
\label{fig:pureSgConvergence}
\end{figure}

Fig. \ref{fig:pureSgConvergence} plots the primal and dual
function values\footnote{The primal value is computed by
$\sum_{s}U_s(x_s(k))$. For the dual value, we have two different
cases. Given the dual variable $\lambda(k)$ at time $k$, if the
global min-cost tree is computed (approximately) at time $k$, then
the dual value is given by $\theta(\lambda(k))$; if the local
min-cost tree is computed at time $k$, then the dual value is
given by $\theta^{(q)}(\lambda(k))$.} and session rates versus the
subgradient iterations for profiles $A1$-$A3$.\footnote{We omit
the figures for profiles $B1$-$B3$ and $C1$-$C3$ since they just
show similar convergence results. Actually, they have even better
convergence results than those of $A1$-$A3$.} It shows that the
rate allocation converges as the primal and dual function values
converge. In some test cases, we have experienced small
oscillation in the allocated rates even though the primal and dual
function values converge. The time-average rates converge in all
cases.


\begin{figure}[t]
\begin{center}
\begin{minipage}{2in}
\begin{center}
\includegraphics[width=2in]{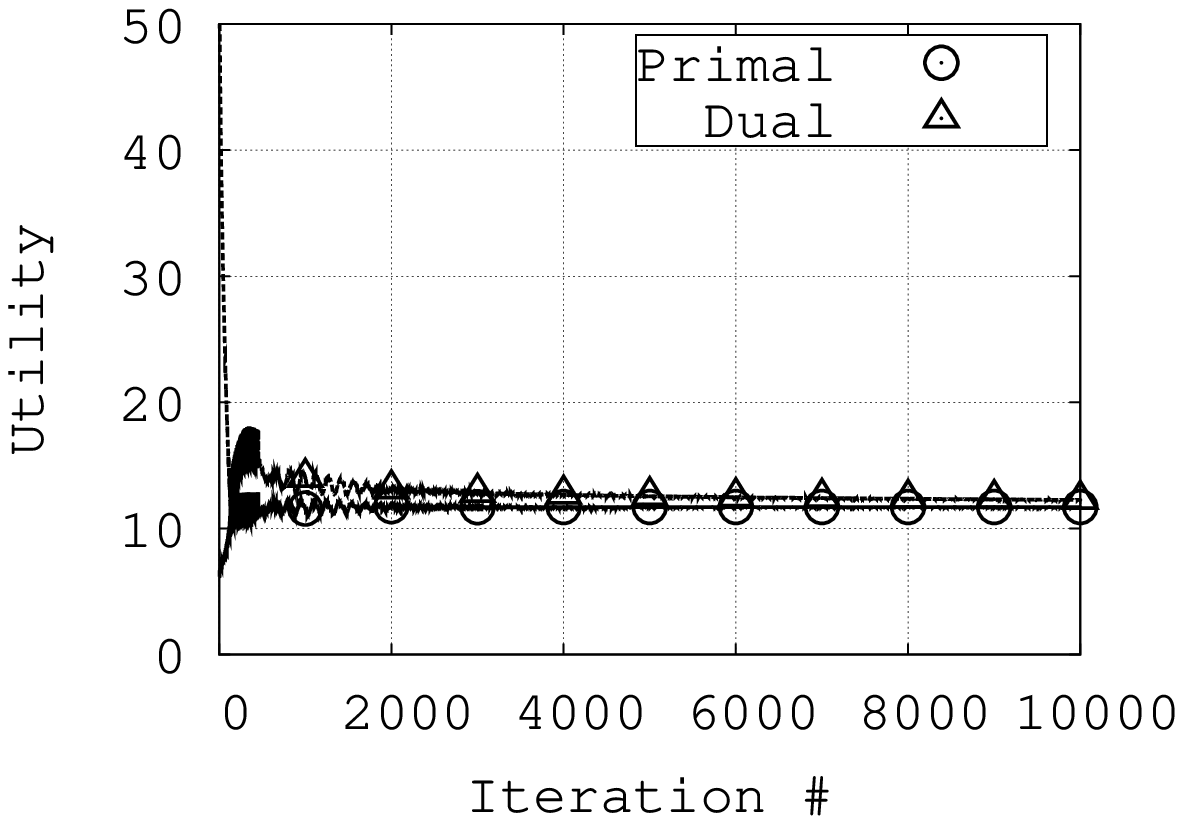} \\ (a) Objective values of A1
\label{fig:A1colGen5obj}
\end{center}
\end{minipage}
\begin{minipage}{2in}
\begin{center}
\includegraphics[width=2in]{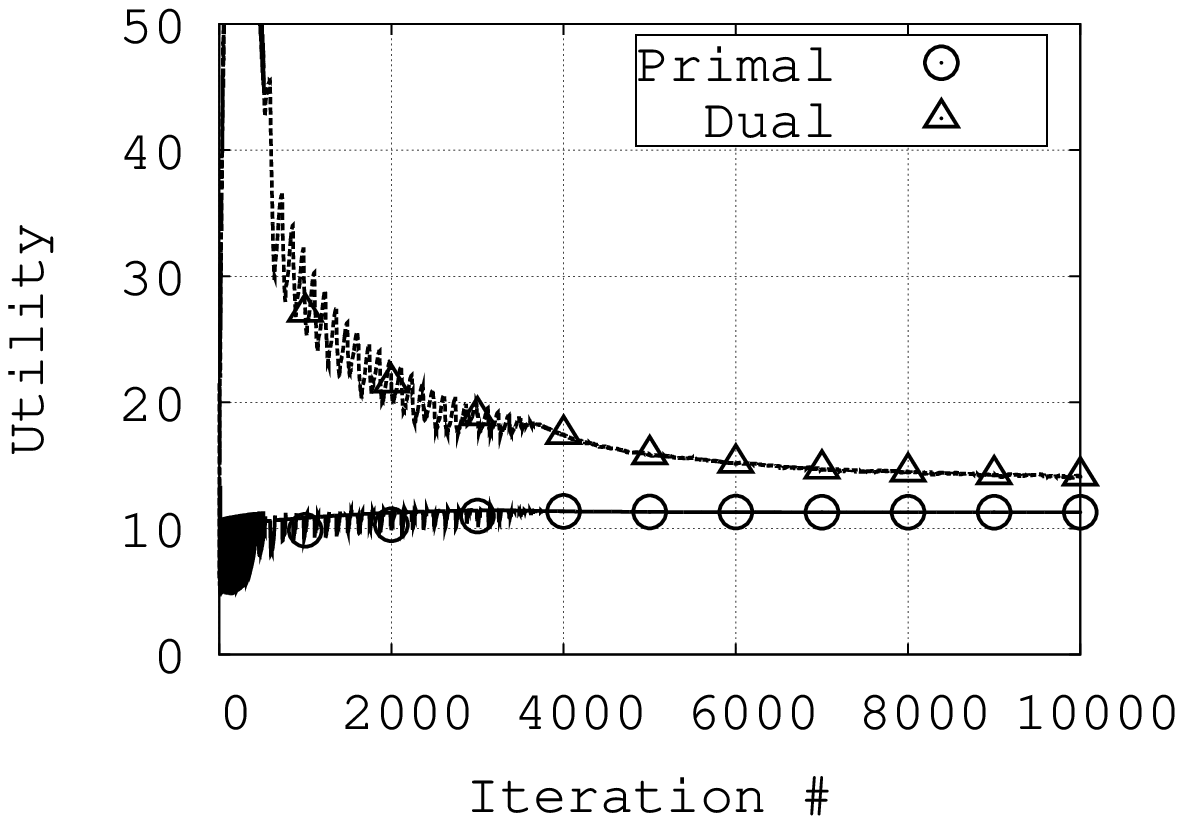} \\ (b) Objective values of A2
\label{fig:A2colGen5obj}
\end{center}
\end{minipage}
\begin{minipage}{2in}
\begin{center}
\includegraphics[width=2in]{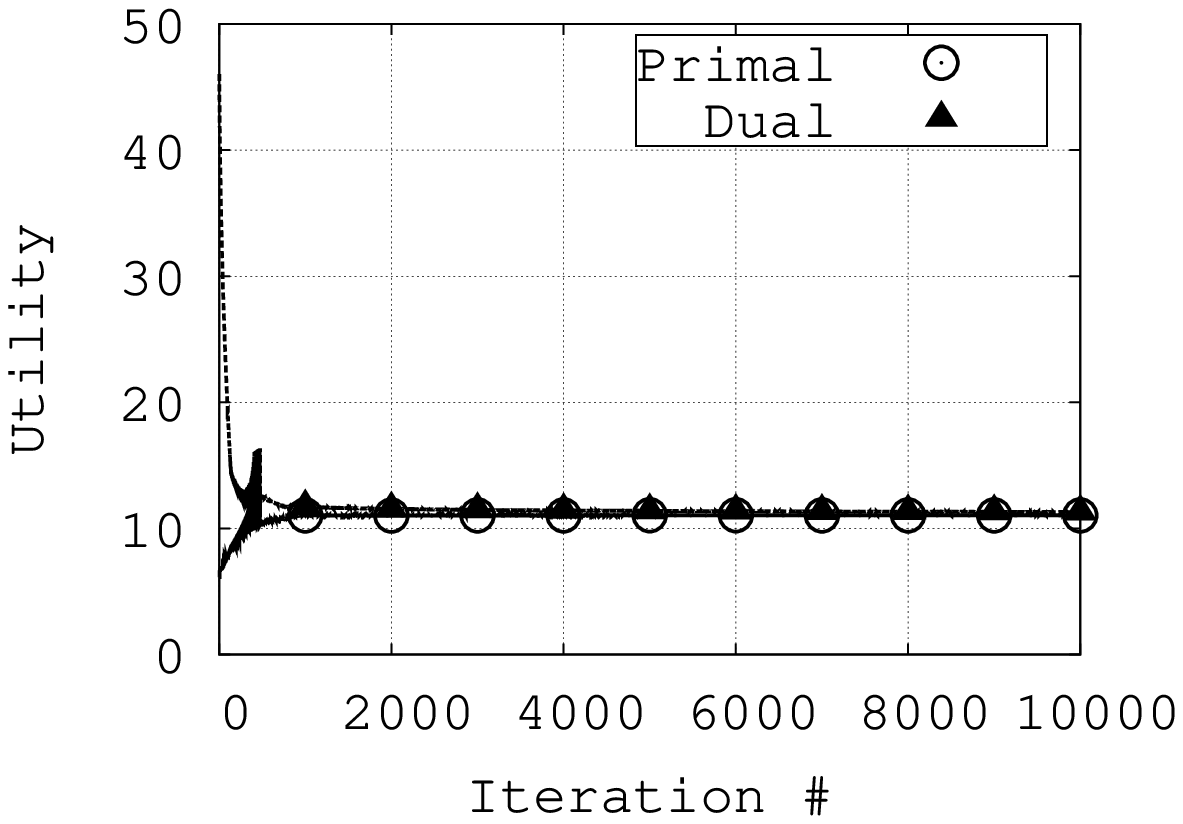} \\ (c) Objective values of A3
\label{fig:A3colGen5obj}
\end{center}
\end{minipage}
\end{center}
\begin{center}
\begin{minipage}{2in}
\begin{center}
\includegraphics[width=2in]{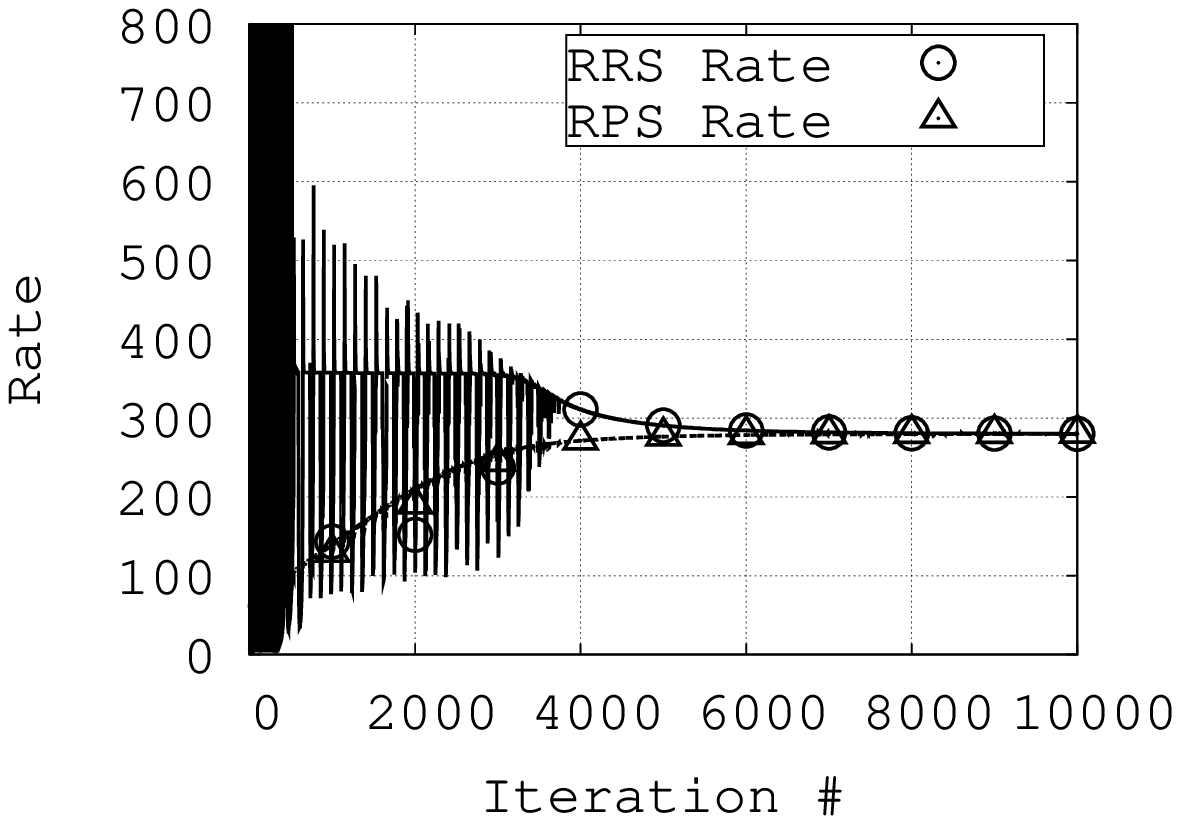} \\ (d) Rates of A2
\label{fig:A2colGen5rates}
\end{center}
\end{minipage}
\begin{minipage}{2in}
\begin{center}
\includegraphics[width=2in]{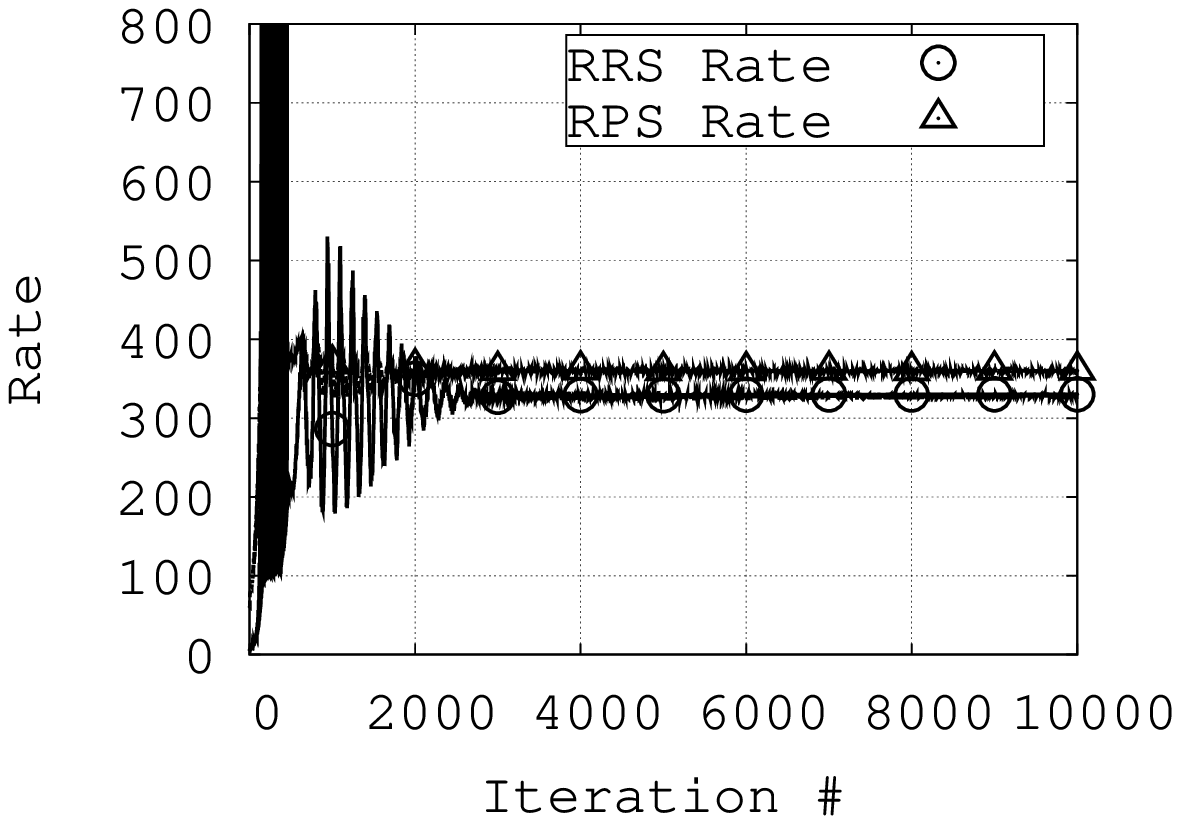} \\ (e) Rates of A1
\label{fig:A1colGen5rates}
\end{center}
\end{minipage}
\begin{minipage}{2in}
\begin{center}
\includegraphics[width=2in]{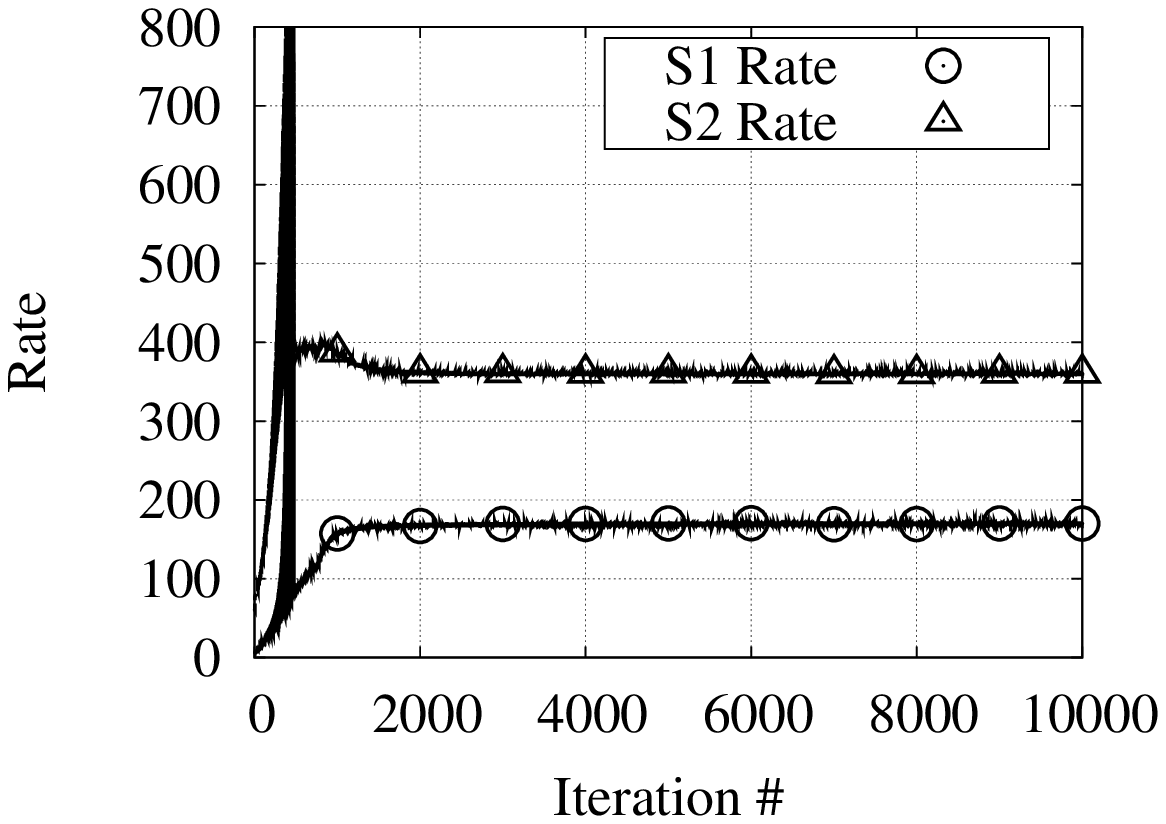} \\ (f) Rates of A3
\label{fig:A3colGen5rates}
\end{center}
\end{minipage}
\end{center}
\caption{Convergence of the subgradient algorithm with column generation; $\Delta = 5$.}
\label{fig:SgColGen5Convergence}
\end{figure}

\begin{figure}[t]
\begin{center}
\begin{minipage}{2in}
\begin{center}
\includegraphics[width=2in]{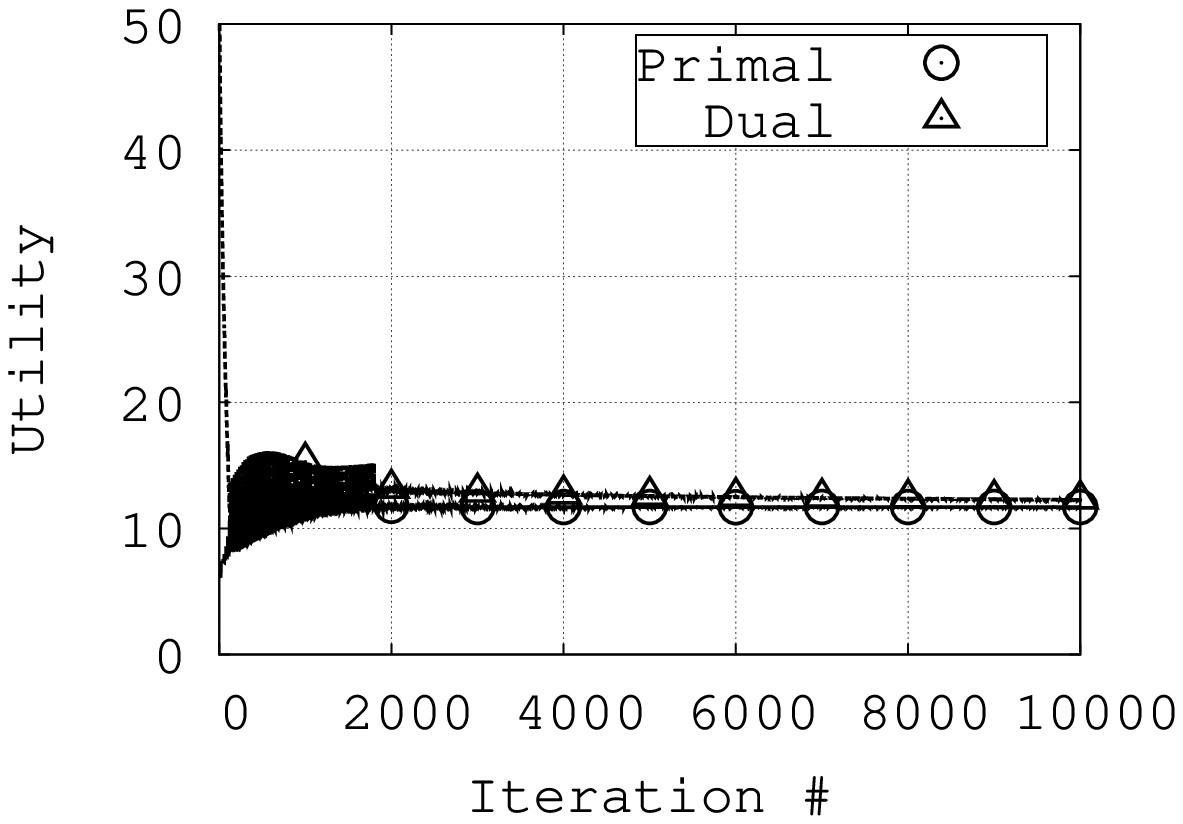} \\ (a) Objective values of A1
\label{fig:A1colGen20obj}
\end{center}
\end{minipage}
\begin{minipage}{2in}
\begin{center}
\includegraphics[width=2in]{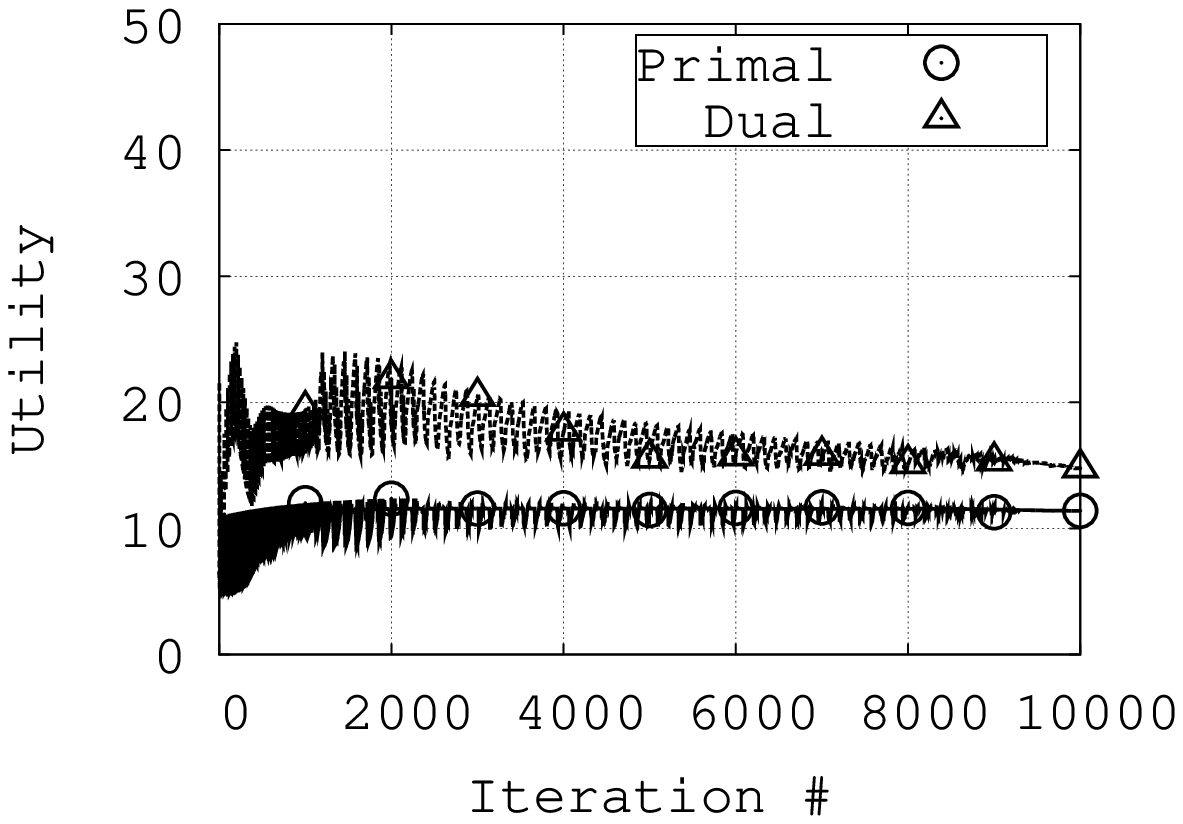} \\ (b) Objective values of A2
\label{fig:A2colGen20obj}
\end{center}
\end{minipage}
\begin{minipage}{2in}
\begin{center}
\includegraphics[width=2in]{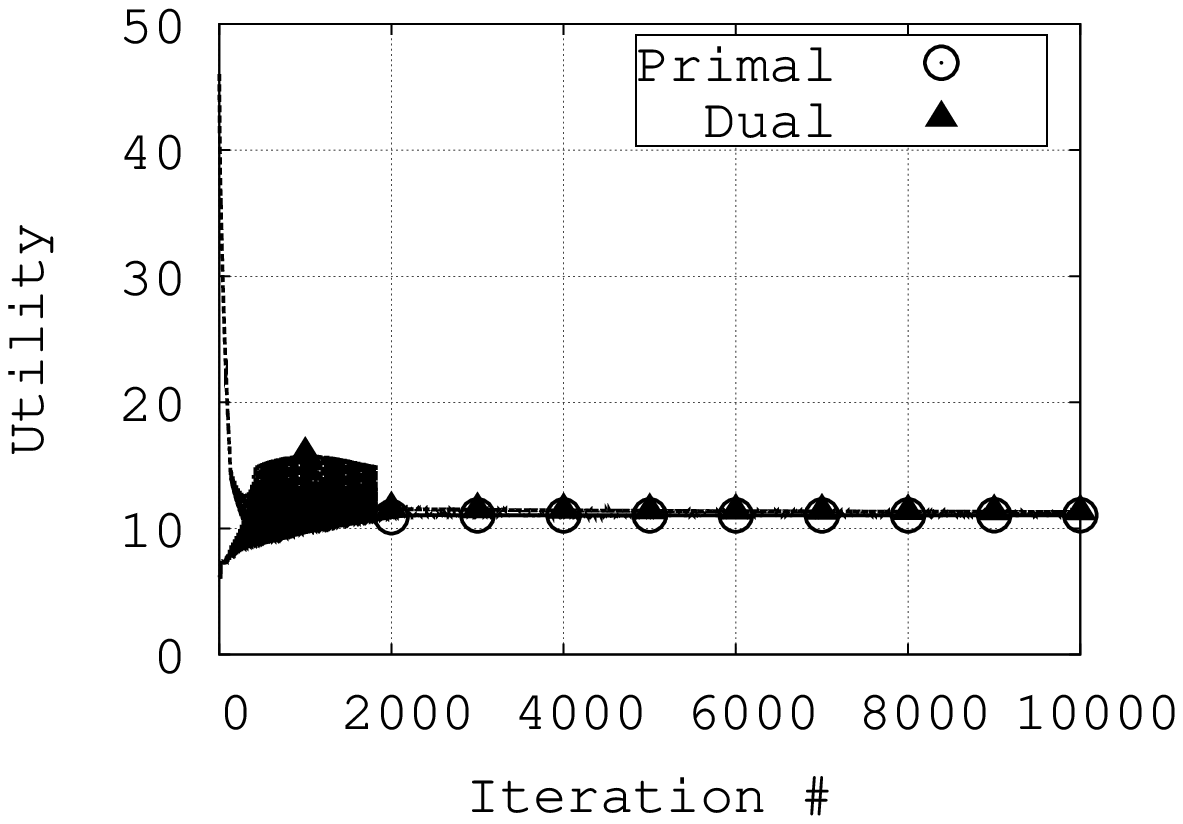} \\ (c) Objective values of A3
\label{fig:A3colGen20obj}
\end{center}
\end{minipage}
\end{center}
\begin{center}
\begin{minipage}{2in}
\begin{center}
\includegraphics[width=2in]{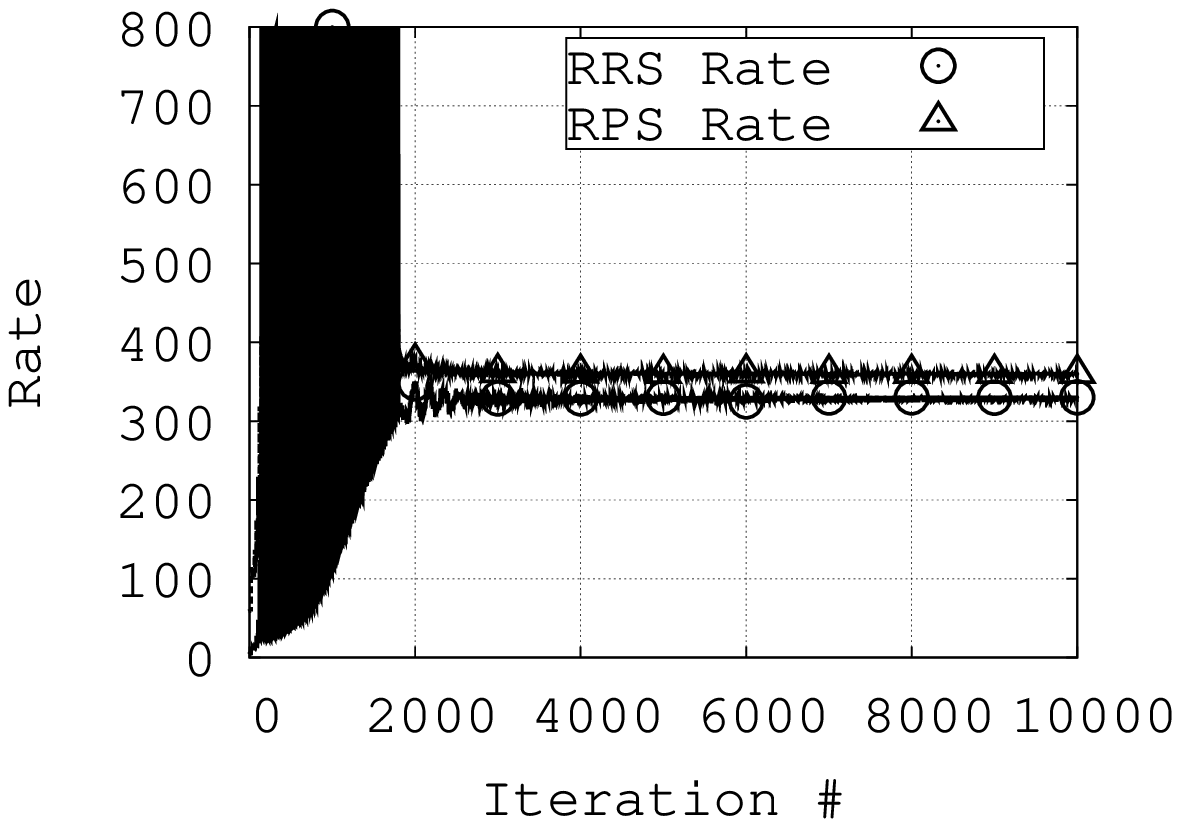} \\ (d) Rates of A1
\label{fig:A1colGen20rates}
\end{center}
\end{minipage}
\begin{minipage}{2in}
\begin{center}
\includegraphics[width=2in]{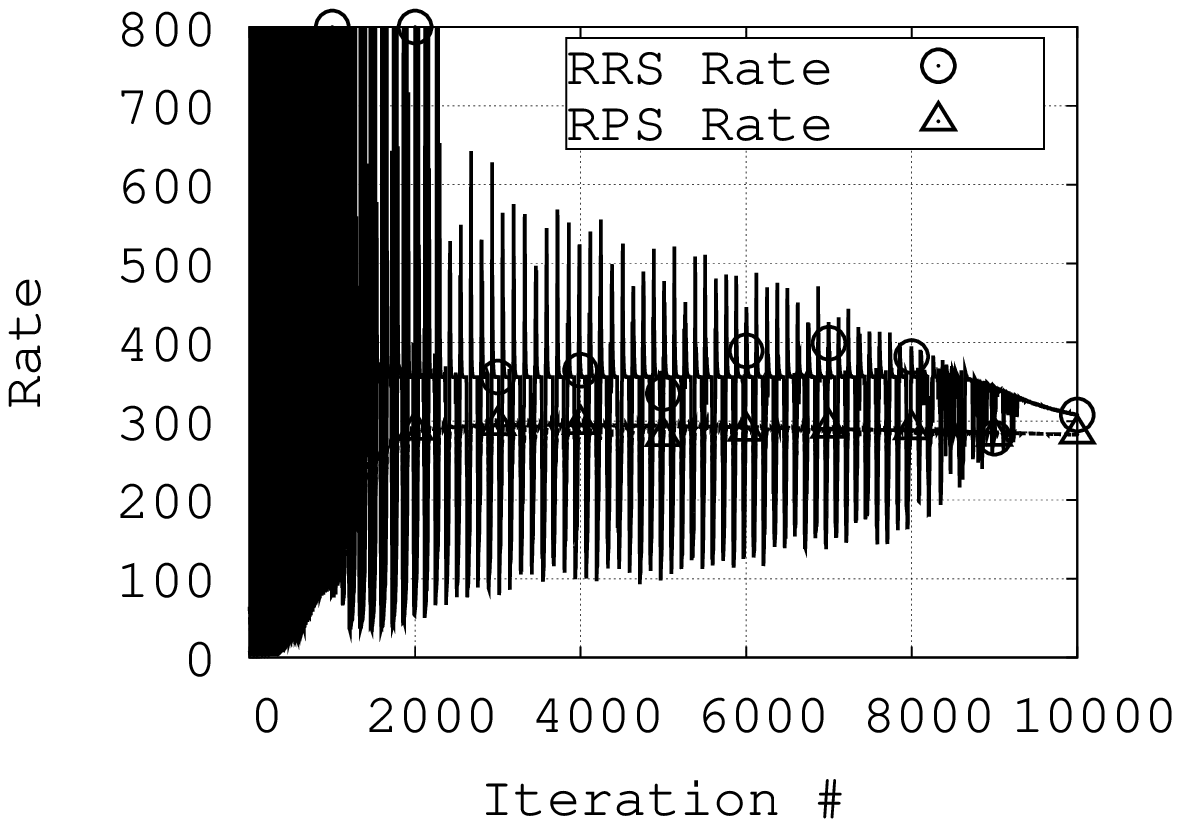} \\ (e) Rates of A2
\label{fig:A2colGen20rates}
\end{center}
\end{minipage}
\begin{minipage}{2in}
\begin{center}
\includegraphics[width=2in]{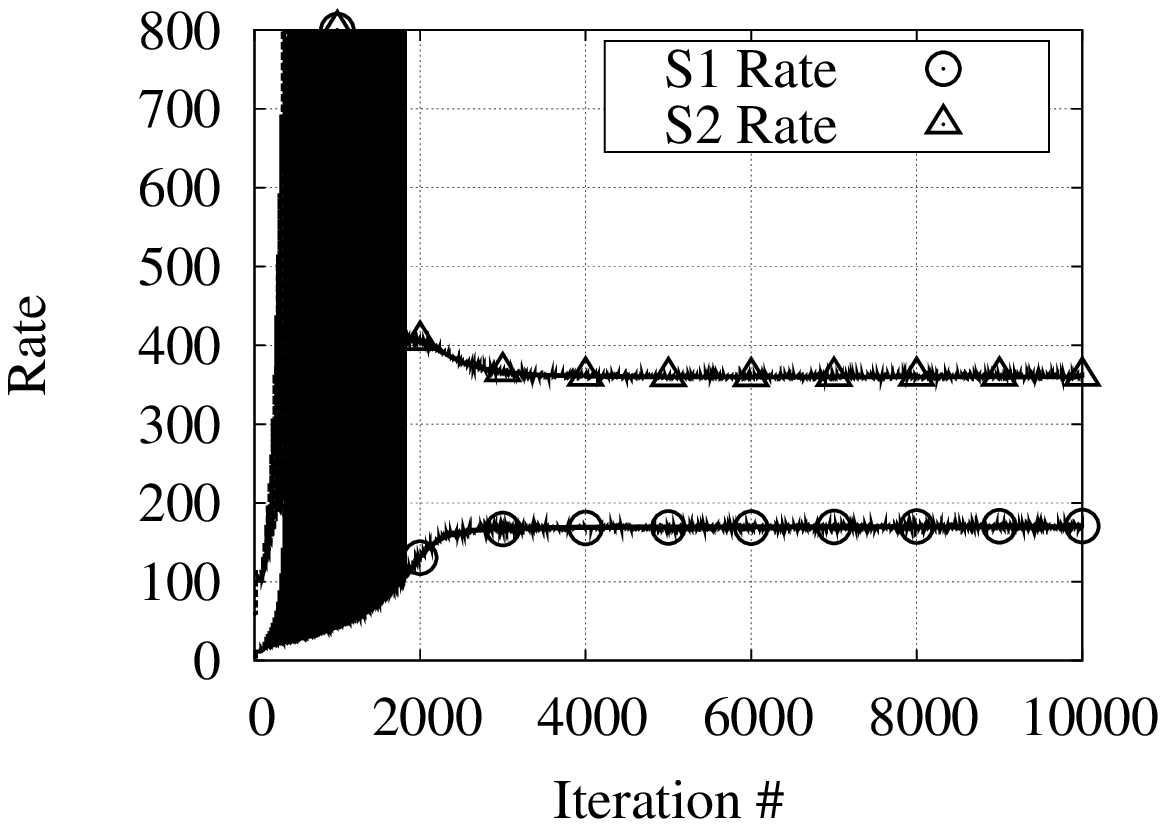} \\ (f) Rates of A3
\label{fig:A3colGen20rates}
\end{center}
\end{minipage}
\end{center}
\caption{Convergence of the subgradient algorithm with column generation; $\Delta = 20$.}
\label{fig:SgColGen20Convergence}
\end{figure}

\begin{figure}[t]
  \centering
  \includegraphics[width=2in]{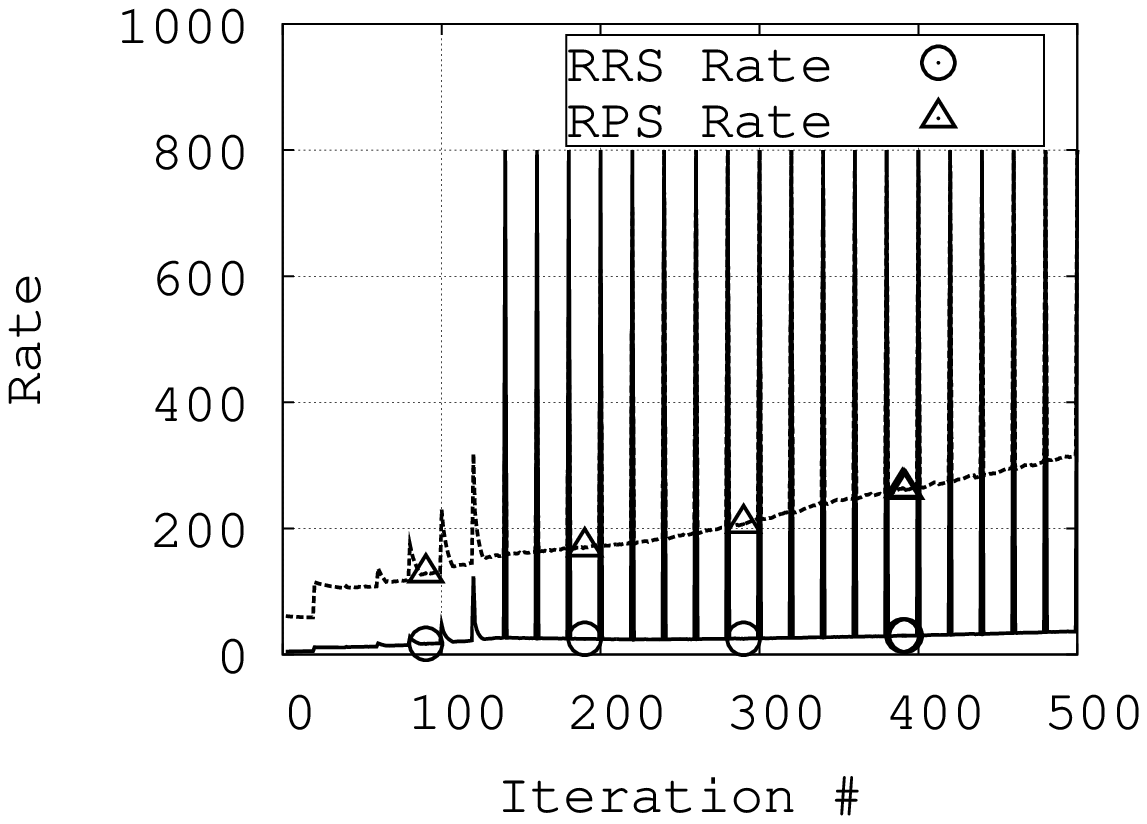}
  \caption{Rates of profile A1 on the initial 500 iterations with the column generation method; $\Delta = 20$.}
  \label{fig:SgColGen20ConvergenceInitialZoom}
\end{figure}

We have also tested the column generation method with the same
profiles. Fig. \ref{fig:SgColGen5Convergence} and Fig.
\ref{fig:SgColGen20Convergence} show the convergence of the primal
and dual function values and the rate allocation when $\Delta = 5$
and $20$, respectively. Note that there is a trade-off in
selecting the size of $\Delta$ (see also Remark 1 after Algorithm
\ref{algo:colgen_imperfect}). As $\Delta$ increases, the total
number of iterations needed for the final convergence (within a
margin) tends to increase while the number of global min-cost tree
computation decreases. Therefore, if the global min-cost tree
computation dominates the overall cost and time complexity, then
we should use a large $\Delta$; on the other hand, if the message
communication required in the iterations dominates the overall
cost and time, we should use a small $\Delta$. There is a clear
benefit of using the column generation method (with $\Delta > 1$)
when the message communication cost is relatively low.

Moreover, the total number of iterations needed for convergence
does not seem to increase as fast as $\Delta$ increases. For
example, with the pure subgradient method ($\Delta = 1$) in
profile $A3$, the convergence takes place at around iteration
$1000$, as shown in Fig. \ref{fig:pureSgConvergence}(f).
On the other hand,
with $\Delta = 5$ or $20$, the convergence takes place at around
iteration $2000$ or $3000$, respectively, as shown in Fig.
\ref{fig:SgColGen5Convergence}(f) and Fig. \ref{fig:SgColGen20Convergence}(f).
In contrast, the numbers of global min-cost tree computation for
those three cases are $1000$, $400$, and $150$, respectively. Note
that profile $A3$ is the worst among the nine test cases in terms
of how the convergence speed of the column generation method
compares to that of the pure subgradient method. In some profiles,
the column generation method shows even faster convergence in
iteration numbers. Therefore, the column generation method can be
quite attractive.

The figures show that there are peaks when a new tree is
introduced after a step of global min-cost tree computation. This
is because the dual cost of the newly selected tree is quite low
at the moment. Even though there are such peaks, the algorithm
quickly adjusts the rates. Fig.
\ref{fig:SgColGen20ConvergenceInitialZoom} shows the rate
allocation during the initial $500$ iterations with $\Delta = 20$
for the column generation method. For every $20$ iterations, there
is a peak; but the rate is quickly adjusted right after the peak.
Such peaks disappear when the algorithm no longer introduces new
trees. Note that this does not mean that all trees have been
introduced. It means that the set of already-introduced trees is
sufficient for the algorithm to achieve the optimal rate
allocation.

\begin{table}[t]
\caption{The number of selected trees (the value of $q$)}
\label{tab:value_of_q}
\begin{center}
\begin{tabular}{|c|c|c|c|}
\hline
Profile & Large session & Small session & q \\
\hline
A1 & 145 & 102 & 247 \\
A2 & 393 & 102 & 495 \\
A3 & 91 & 93 & 184 \\
\hline
B1 & 51 & 53 & 104 \\
B2 & 51 & 53 & 104 \\
B3 & 51 & 53 & 104 \\
\hline
C1 & 102 & 49 & 151 \\
C2 & 102 & 102 & 204 \\
C3 & 102 & 11 & 113 \\
\hline
\end{tabular}
\end{center}

\end{table}

We next discuss the magnitude of the value $q$ in our simulation. Recall that $q$ is the number of trees that have been selected as an (approximate) min-cost tree at some time. In our test cases, most of the values of $q$ are between $104$ and $247$, and the largest value is $495$ in profile A2 (see Table  \ref{tab:value_of_q}).
Note that the number of candidate trees grows extremely rapidly as the network size increases.
In our simulation, the nodes of each session form a complete (overlay) graph. For a complete graph, the number of possible trees is known to be $(|R_s|+1)^{|R_s|-1}$ where $|R_s|$ is the number of receivers of a session $s$ \cite{cui2004maxmin}.
Then, even without considering out-of-session nodes, a large session with $90$ receivers may have $91^{89}$ candidate trees and a small session with $10$ receivers may have $11^{9}$ trees. We see that the number of trees used in the algorithm is relatively small.

\subsection{Event-driven simulation}

In order to evaluate the performance of the algorithms more carefully, we develop an event-driven simulator to trace the behavior of the control packets and measure the real queue sizes.

We first outline a design of the signaling/control protocol.
A signaling packet contains a list of link IDs (32 bits each), a 32-bit virtual source rate
and a 32-bit multicast session ID.
On each time slot, one signaling
packet will be sent towards each node on the multicast tree selected by the algorithm.
Feedback packets are used to propagate the dual costs of
the outgoing links from each node back to each source.
A feedback packet contains the IDs and dual costs (32 bits each) of the node's outgoing links
and the 32-bit node ID.
Every signaling/feedback packet has an additional 20-byte header containing miscellaneous information.
Hence, in typical settings, the size of a control packet is no more than $1000$ bytes.

We calculate the algorithm overhead due to the control packets.
For a network with $n$ nodes and $m$ links, on each time slot,
the total size of all the signaling packets from a source is at most $(20 \times 8 + 32 + 32) n + 32 m$ bits, and the total size of all the feedback packets received by a source
is $(20 \times 8 + 32)n + (32+32)m$ bits.
Consider a network with $300$ nodes and $3000$ links and suppose the time slot size is $0.5$ seconds.
On each time slot, the total size of the forward signaling packets is $163,200$ bits (corresponding to a control traffic rate of $326.4 $Kbps), and the total size of the feedback packets is $249,600$ bits (corresponding to a control traffic rate of $499.2$ Kbps).
If the source data rate is $100$ Mbps, the overhead is only $0.8256 \%$. Furthermore, the control traffic rate is independent of the source data rate, which means the larger the source rate is,
the smaller the overhead is.

The control packets are routed on the shortest path, measured by the hop count and
are transmitted at a higher priority than the data packets.
An event is triggered to update the link flow rate or the link dual cost when a control packet
arrives at its destination, which is either a source or a network node.

To illustrate the algorithm updates and the exchange of the control packets, we present the pseudo-code of the algorithm (see Algorithm \ref{algo:pseudo}).
Due to the propagation delays of the signaling/feedback packets, the information about the tree flow rates and network costs is generally not consistent across the sources and nodes at each point in time. Let $r_e$ be the link flow rate known at each link $e$ and let
$\tilde{\lambda}_{s, e}$ be link $e$'s dual cost known by source $s$.

\begin{algorithm}
\caption{Pseudo-code: Column Generation with Imperfect Global Tree Scheduling}
\label{algo:pseudo}
{\bf Actions on each time slot:}
\beit
\item At each link $e$:
\beit
\item update link dual cost by $\lambda_e \leftarrow [\lambda_e - \delta_e (c_e - r_e)]_+$
\item Reset link flow rate $r_e \leftarrow 0$
\eeit
\item At each source $s$:
\beit
\item
under the current dual cost $\tilde{\lambda}_s$, either
(a) search for a local min-cost tree from the current collection of trees,
or (b) compute an approximate min-cost tree. If the approximate min-cost tree
is not in the current collection of trees, introduce it into the current collection. Denote the local/approximate min-cost tree by $t(s, \tilde{\lambda}_s)$
\item
update
the source rate $x_s$ based on the dual cost of tree $t(s, \tilde{\lambda}_s)$
according to (\ref{eq:sub_x})
\item
source $s$ sends one signaling packet towards each node on the multicast tree
$t(s, \tilde{\lambda}_s)$
\eeit
\item At each node $n$:
\beit
\item node $n$ sends one feedback packet towards each source of the multicast sessions
\eeit
\eeit
{\bf Actions when a feedback packet arrives at source $s$}:
source $s$ updates the relevant link dual cost by $\tilde{\lambda}_{s, e} \leftarrow \lambda_e$

{\bf Actions when a signaling packet arrives at node $n$}:
node $n$ updates link flow rate $r_e$
by $r_e \leftarrow r_e + x_s$ if link $e$ is on the multicast tree (for $s$) indicated by the signaling packet
\end{algorithm}

We trace the behavior of the real data queues at the burst level instead of the packet level to reduce the simulation time.
On each time slot, for each link, the amount of data a link can transmit is calculated, which is the difference of the link capacity and the amount of control packets transmitted
during that time slot. Then, the burst of data is pushed to the next hop. 


We simulate our algorithm on a commercial ISP network topology
obtained from the Rocketfuel project \cite{RocketFuel}
consisting of $295$ nodes and $1086$ links, and one multicast session with $39$ receivers.
We assign $5$ Gbps link capacity to each of the critical links and $1$ Gbps to each of the other links, 
The signaling/feedback packet size is 400 bytes for our experiments.
The link propagation delay is $100$ ms.
In order to take into account the tree computation time, a tree computed on time slot $k$ will be used
to transmit data on time slot $k + 1$. The time slot size is $1$ second.
the algorithm computes a global approximate tree every $10$ seconds.

\begin{figure}[t]
\begin{center}
\begin{minipage}{2.5in}
\begin{center}
\includegraphics[width=2.5in]{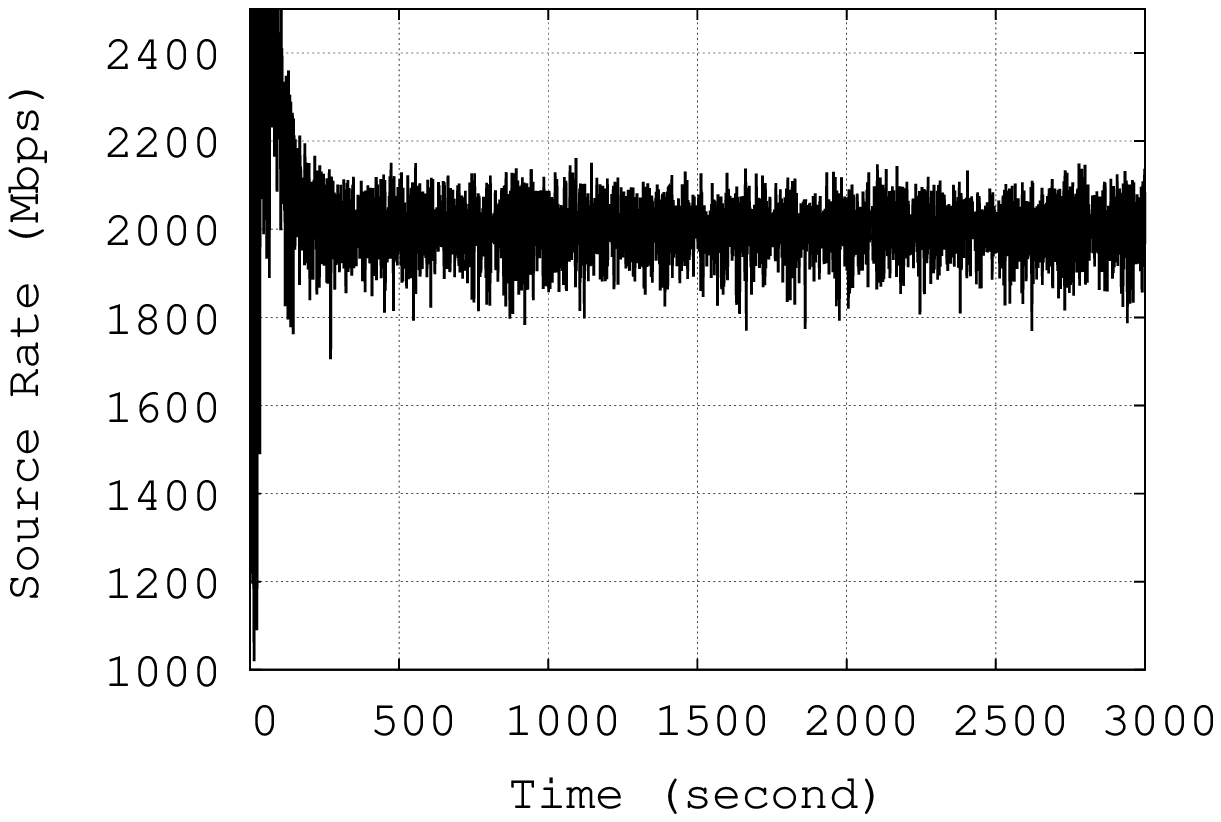} \\ (a) Source rate
\end{center}
\end{minipage}
\begin{minipage}{2.5in}
\begin{center}
\includegraphics[width=2.5in]{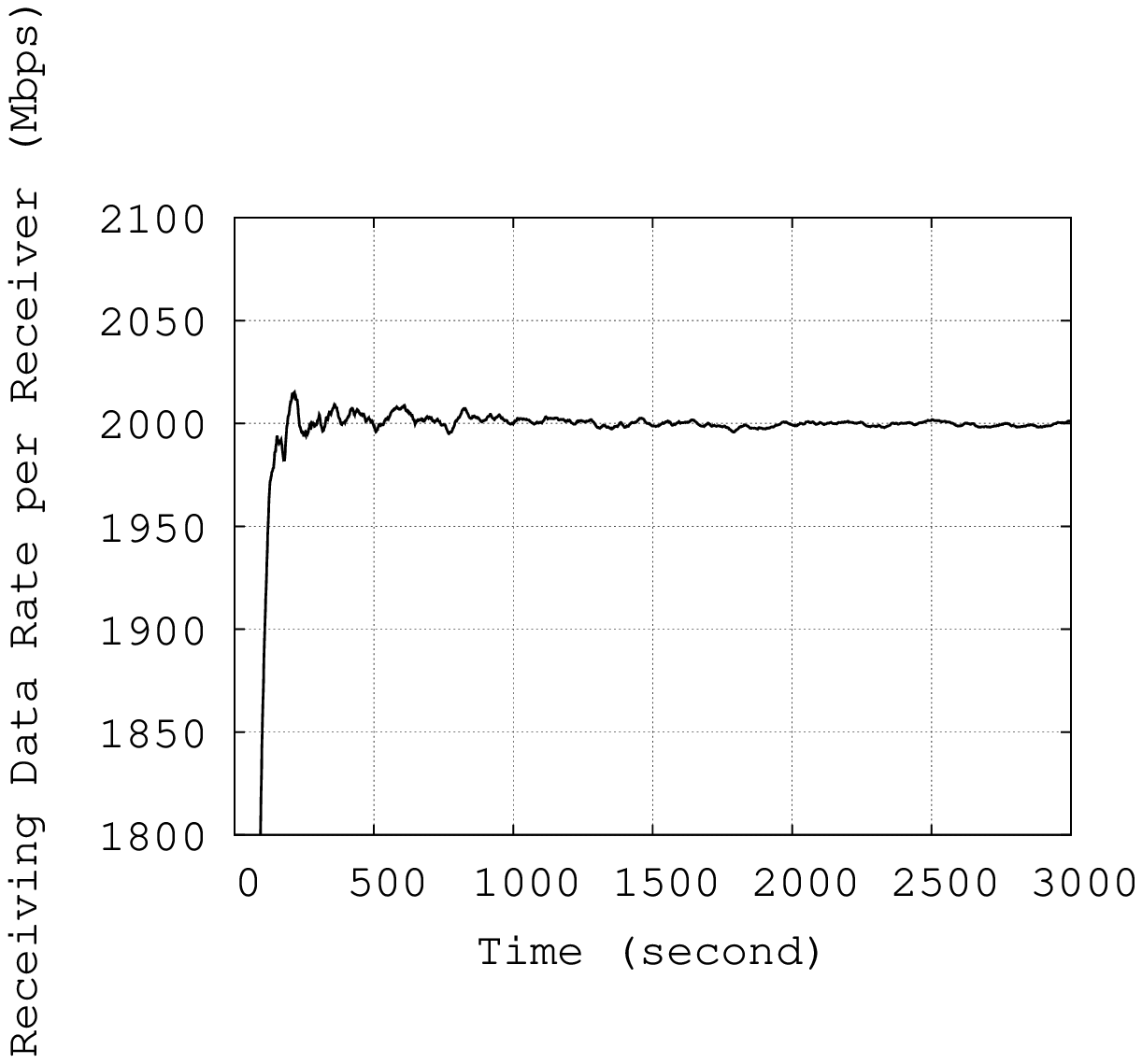} \\ (b) Average receiving rates
\end{center}
\end{minipage}
\begin{minipage}{2.5in}
\begin{center}
\includegraphics[width=2.5in]{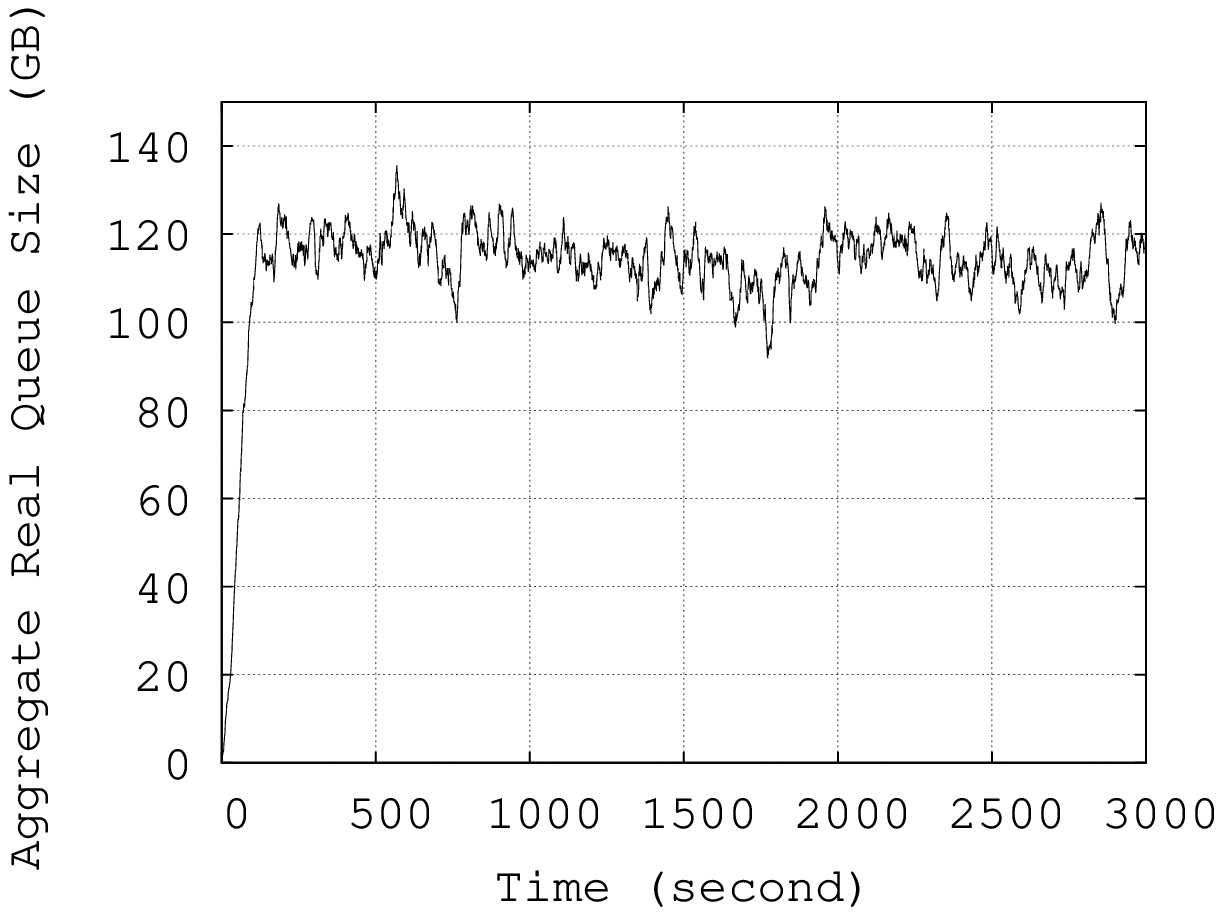} \\ (c) Aggregate real queue backlog
\end{center}
\end{minipage}
\begin{minipage}{2.5in}
\begin{center}
\includegraphics[width=2.5in]{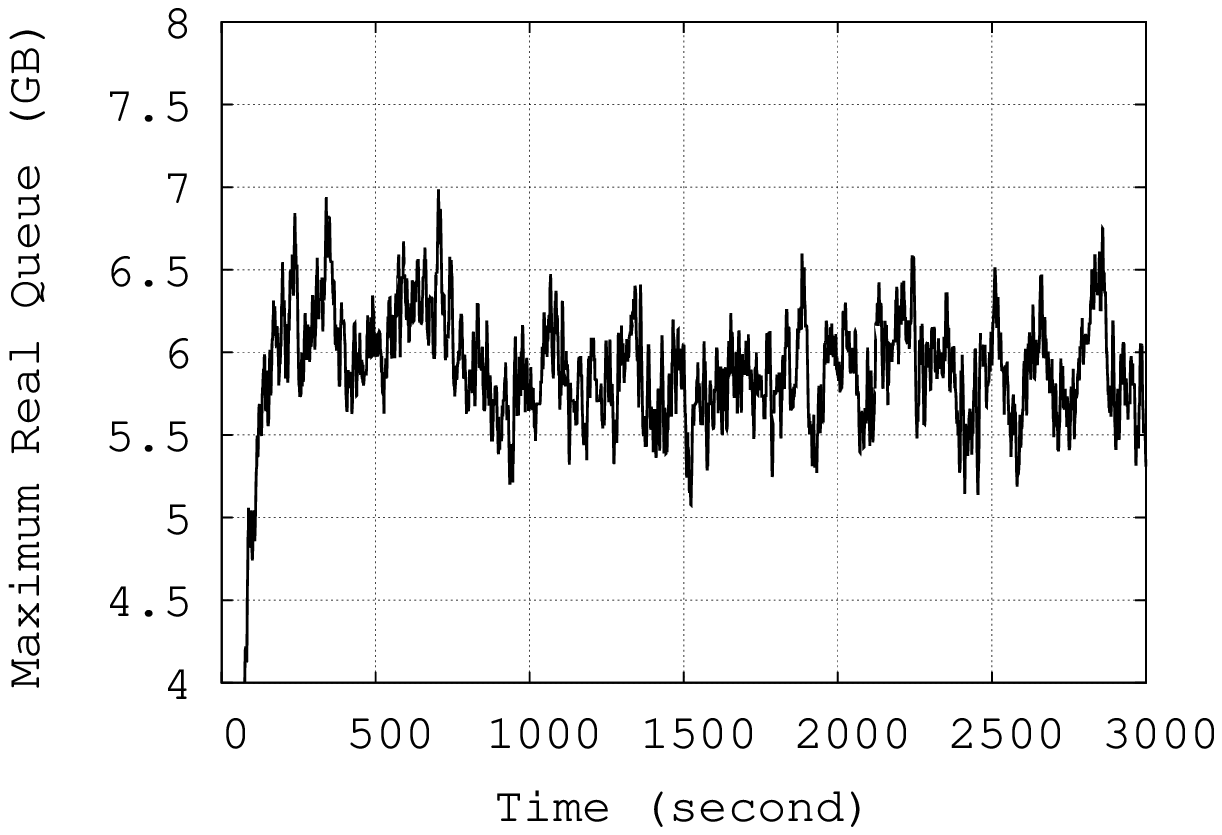} \\ (d) Maximum real queue backlog
\end{center}
\end{minipage}
\end{center}
\caption{Algorithm performance on a commercial ISP topology; $\Delta = 10$.}
\label{fig:large}
\end{figure}

Fig. \ref{fig:large}(a) shows the evolution of the source rate, and Fig. \ref{fig:large}(b) shows
the evolution of the average receiving rate at the receivers.
We see that the algorithm works well under fairly realistic conditions, including asynchronous exchange of the control information, substantial link
propagation delays and tree computation delay.
Fig. \ref{fig:large}(c) and (d) show the aggregate real queue size over the entire network
and the maximum real queue size across the links, respectively. The average queue size per link is calculated to be around $110$ MB (dividing the aggregate queue size by $1086$ links). The maximum real queue size is around $6$ GB, which is large but not prohibitive.

Note that the queue sizes shown in Fig. \ref{fig:large}(c) and (d) are mostly 
built up at the transient phase of the simulation, which is the phase at the beginning of the algorithm operation before it finds the right transmission rates. Once entering the steady state,
the source rates converge to some feasible values and the queue sizes stop growing
but oscillate slightly. The oscillation is due to that a multicast session hops among different multicast trees even in the steady state. As a result, some of the links may be temporarily and slightly overloaded. We see from the simulation results that the magnitude of the oscillation is very small.

The intended content distribution application operates in a relative static network environment with long-lasting sessions. In that case, the buffer requirement can be decided based on the steady-state behavior of the algorithm, since the system will stay in the steady state mostly. The results indicate that the buffer requirement for absorbing the steady-state queue oscillation is very small.

If we want the algorithm to cope well with occasional changes in the environment, including the arrivals or departures of multicast sessions, link failures or additions, changes in link bandwidth, etc., it is desirable to ensure that the transient queue sizes are not too large. In the simulation results, the average queue size per link is less than the smallest bandwidth-delay product, where the delay here means the time slot size. The maximum queue size is not small but not prohibitively large either. In particular, if we apply the algorithm to overlay content distribution, which is the intended scenario of the paper, the nodes are content servers and the buffers are in the content servers. In that case, it is fairly easy and inexpensive to add large buffers (for instance, consider having 6 GB memory at a larger server). Since we are considering a dedicated content distribution network, which is owned by a single provider and is not shared with other applications, delay due to long queues is not an issue for the intended application (mostly file transfers). Finally, the transient queue sizes depend on the convergence speed of the algorithm, which in turn depends critically on the parameter $\delta$. They can potentially be made smaller if we tune the parameter $\delta$ appropriately.

\section{Related works}
\label{section:related}

We now briefly discuss additional related work. The work in \cite{ZhengCX09} focuses on the queueing process when the universal swarming technique
is applied to content delivery. It investigates the stability of the queues under the proposed algorithms and the analytical tool is the Lyapunov drift analysis about Markov processes. In this paper, we have an deterministic optimization problem as opposed to a stochastic stability problem and the analytical tool is mostly convex optimization.
The work in \cite{CGY00} aims at minimizing the
link congestion (equivalently, maximizing the throughput)
for multiple multicast sessions.
Its authors propose a heuristic centralized
tree packing algorithm, where each tree for
each session is computed by using cutting-plane
inequalities and the branch-and-cut algorithm.
The authors of
\cite{Jansen02anapproximation} study the
multicast congestion problem, where a single multicast tree is used
for each multicast session. They
present a centralized approximation
algorithm to minimize the maximum link congestion.
\cite{WCJ04, CHLCD07,
Lun05achievingminimum-cost} all use the technique of network coding, which can achieve
the multicast network capacity.
In \cite{WCJ04}, the authors compare
network coding and tree packing, and show by
simulation that
tree packing performs comparably to network
coding in terms of throughput.
Both
\cite{CHLCD07} and \cite{Lun05achievingminimum-cost}
present distributed algorithms to compute
the optimal rate allocation for
network coding based multicast.
However, in order to achieve the
multicast capacity, appropriate encoding
and decoding functions are needed, which are difficult to find. A survey of optimization problems in multicast routing
can be found in \cite{OliveP05}.

The authors of \cite{Bin06,
ZNTP07, KR06} model and analyze peer-assisted file
distribution systems.
In \cite{Bin06}, in order to decrease the cross-ISP traffic
generated by BitTorrent,
peers are encouraged to receive file chunks from peers within the same ISP.
In \cite{ZNTP07}, the authors study how to accelerate
data transmission by opening multiple connections
on multiple paths.
In \cite{KR06}, the minimum data distribution time
is derived for a peer-assisted network where
the bottlenecks are at the access links.

The multipath routing problem has been
studied in \cite{KMT07, Paganini06, LinS06, LestasV04}.
The work in \cite{KMT07} studies congestion control for multi-path unicast data transfer. It shows that, when the
users are allowed to change their routes,
simple path selection polices of shifting to the paths
with a higher net benefit can lead to utility maximization.
In \cite{Paganini06}, backpressure algorithms
are proposed for congestion control in
a multi-path routing setting.
In \cite{LinS06}, the authors study multi-path utility maximization problems and
develop a distributed algorithm that is amenable
to online implementation.
The work in \cite{LestasV04} solves the problem of maximizing an aggregate utility when all possible paths from the sources to the destinations can be used.

Various systems have been proposed for tree-based live streaming or content delivery (CDN), such as FastReplica \cite{LeeV07} and SplitStream
\cite{Splitstream}.
In FastReplica, the distribution of a file takes a two-phase hierarchical approach. Its performance can be poor when the bottleneck is at the internal of the network \cite{ZCXinfocom08}.
In Splitstream, a file is split into a number of chunks and each chunk uses a separate multicast tree. The main focus of that study is on load balancing of the nodes and improving the resilience to node failures. The performance optimality has not been shown.
A commercial hybrid CDN-P2P streaming system,
{\em LiveSky}, has provided live streaming
for several large-scale events with more than ten
million online users \cite{Yin2009}.
LiveSky adopts the tree-based CDN infrastructure
to transmit chunks from the source to the cache servers.
Later, the end users can receive data from the edge cache
servers, and a mesh-based P2P technique
is used to speed up data sharing among the end users.
It is said that multiple trees are used in the system to transmit chunks
from the data source to the cache servers. However, the details have not been published.


\section{conclusion}
\label{sec:conclusion}

This paper studies the method of universal swarming for content
distribution, which allows multiple sessions to help each other
and improve the overall distribution performance. For relatively
static, infrastructure-based content distribution, we can model
universal swarming as distribution over multiple multicast trees.
That is, the data of each session is distributed by a set of
multicast trees rooted at the source and spanning all the
receivers. Each multicast tree is in general a Steiner tree
containing out-of-session nodes. The question is how to optimally
allocate rates to the multicast trees so that the sum of all
sessions' utilities is maximized. We develop a distributed
subgradient algorithm. Due to the partial linearity of the
problem, there is no standard convergence result for the algorithm
and the algorithm does not converge in the normal sense. We prove
that the subgradient algorithm converges in the time-average
sense. Furthermore, the subgradient algorithm involves an NP-hard
subproblem of finding a min-cost Steiner tree. We adopt a column
generation method with imperfect min-cost tree scheduling. If the
imperfect min-cost tree has an approximation ratio $\rho$, then our
overall utility-optimization algorithm converges to a sub-optimum
with an approximation ratio at least as good as $1/\rho$.


\section{Appendix A: Proofs in Section \ref{sec:subgradient}}

\subsection{Proof of Lemma \ref{lemma:sub}}

\noindent $(a)$
The problem (\ref{formula:primal}) is maximizing a
concave function with linear constraints, the strong duality holds
for (\ref{formula:primal}) and there is no duality gap at the optimum
of (\ref{formula:primal}), i.e.,
$f^* = \theta(\lambda^*)$ for any $\lambda^* \in \Lambda^*$.

\noindent $(b)$
Under fixed $\lambda$, let $\lambda_t = \sum_{e \in t} \lambda_e,
\forall t \in T$.
Define $g_s(y) = U_s(\sum_{t \in T_s} y_t) - \sum_{t \in T_s} y_t \lambda_t$.
For each source $s$, the Lagrangian maximization sub-problem is
\bea
\label{formula:lagrangian_maxmization_1}
& \max \ U_s(x_s) - \sum_{t \in T_s} y_t \lambda_t & \\
\mbox{s.t.} & x_s = \sum_{t \in T_s} y_t & \nonumber \\
& m_s \leq x_s \leq M_s & \nonumber \\
& y_t \geq 0, & \forall t \in T_s, \nonumber
\eea
which is equivalent to
\bea
\label{formula:lagrangian_maxmization_2}
& \max \ g_s(y) & \\
\mbox{s.t.} & m_s \leq \sum_{t \in T_s} y_t \leq M_s & \nonumber \\
& y_t \geq 0, & \forall t \in T_s. \nonumber
\eea
Since $g_s(y)$ is not strictly concave in $y$,
the problem (\ref{formula:lagrangian_maxmization_2}) might have
multiple optimal solutions. We will show that $y^*$ is one of the
optimal solutions, where
\be
y^*_t =
\left\{ \begin{array}{ll}
         [U_s'^{-1} (\gamma(s, \lambda))]_{m_s}^{M_s} & \mbox{if $t = t(s, \lambda)$,}\\
         0 & \mbox{otherwise}. \end{array} \right.
\label{eq:ystarsol}
\ee
Since $g_s(y)$ is a concave function, according to the sufficient optimality
condition, $y^*$ is optimal to
(\ref{formula:lagrangian_maxmization_2}),
if $\nabla g_s(y^*)^T (y - y^*) \leq 0$
for any feasible $y$, where $m_s \leq \sum_{t \in T_s} y_t \leq M_s$ and
$y_t \geq 0, \forall t \in T_s$. Here, $\nabla g_s(y^*)$ is
the gradient column vector and $T$ represents the transpose operation.
Let $z = y - y^*$ be a shorthand, and therefore,
$\nabla g_s(y^*)^T (y - y^*)
= \sum_{t \in T_s} \frac{\partial g_s(y^*)}{\partial y_t} z_t$.

\noindent {\bf Case $1$} $ m_s < y^*_{t(s, \lambda)} < M_s$:
By (\ref{eq:ystarsol}) and the fact that $\lambda_t \geq \gamma(s, \lambda)$ for all $t \in T_s$,
\be
\frac{\partial g_s(y^*)}{\partial y_t} = U'_s(\sum_{t \in T_s} y^*_t) - \lambda_t
\left\{ \begin{array}{ll}
         = 0 & \mbox{if $t = t(s, \lambda)$,}\\
         \leq 0 & \mbox{otherwise}. \end{array} \right.
\nonumber
\ee
Hence, if $t = t(s, \lambda)$, then $\frac{\partial g_s(y^*)}{\partial y_t} z_t = 0$;
if $t \not= t(s,\lambda)$ (and hence, $y^*_t=0$), then $\frac{\partial g_s(y^*)}{\partial y_t} z_t \leq 0$.
Thus, $\nabla g_s(y^*)^T (y - y^*) \leq 0$ for any
feasible $y$.

\noindent {\bf Case $2$} $y^*_{t(s, \lambda)} = m_s$: In this case, we claim
\be
\frac{\partial g_s(y^*)}{\partial y_t}
= U'_s(m_s) - \lambda_t \leq 0, \forall t \in T_s. \nonumber
\ee
To see this,
note that $U'_s$ is a non-increasing function since $U_s$ is concave.
Then, $m_s = y^*_{t(s,\lambda)} = [U_s'^{-1} (\gamma(s, \lambda))]_{m_s}^{M_s}
\geq U_s'^{-1} (\gamma(s, \lambda))$,
and $U'_s(m_s) \leq U'_s(U_s'^{-1} (\gamma(s, \lambda))) = \gamma(s, \lambda)$.
Also, $\lambda_t \geq \gamma(s, \lambda)$ for all $t \in T_s$. Hence, the claim
is true.
For any feasible $y$, $z_t \geq 0$ for all $t \in T_s$.
Therefore, $\nabla g_s(y^*)^T (y - y^*) \leq 0$.

\noindent {\bf Case $3$} $y^*_{t(s, \lambda)} = M_s$: In this case, we claim
\begin{align}
\frac{\partial g_s(y^*)}{\partial y_t}
= U'_s(M_s) - \lambda_t
\left\{ \begin{array}{ll}
         \geq 0 & \mbox{if $t = t(s, \lambda)$,}\\
         \leq \frac{\partial g_s(y^*)}{\partial y_{t(s, \lambda)}}
         & \mbox{otherwise}. \end{array} \right.
\nonumber
\end{align}
To see $\frac{\partial g_s(y^*)}{\partial y_t} \geq 0$ for the case
$t = t(s, \lambda)$,
$M_s = y^*_{t(s,\lambda)} = [U_s'^{-1} (\gamma(s, \lambda))]_{m_s}^{M_s}
\leq U_s'^{-1} (\gamma(s, \lambda))$,
and $U'_s(M_s) \geq U'_s(U_s'^{-1} (\gamma(s, \lambda))) = \gamma(s, \lambda) = \lambda_t$.
For $t \not= t(s,\lambda)$, since $\lambda_t \geq \gamma(s, \lambda)$,
$\frac{\partial g_s(y^*)}{\partial y_t} \leq \frac{\partial g_s(y^*)}{\partial y_{t(s, \lambda)}}$.
Furthermore, for any feasible $z_t$,
$\sum_{t \in T_s} z_t \leq 0$, $z_{t(s, \lambda)} \leq 0$,
and $0 \leq \sum_{t \in T_s, t \neq t(s, \lambda)} z_t \leq - z_{t(s, \lambda)}$.
Also, $z_t \geq 0$ for $t \not= t(s,\lambda)$.
Hence,
\begin{align}
& \nabla g_s(y^*)^T (y - y^*) \nonumber \\
= & \sum_{t \in T_s, t \neq t(s, \lambda)} \frac{\partial g_s(y^*)}{\partial y_t} z_t
+ \frac{\partial g_s(y^*)}{\partial y_{t(s, \lambda)}} z_{t(s, \lambda)} \nonumber \\
\leq & \sum_{t \in T_s, t \neq t(s, \lambda)} \frac{\partial g_s(y^*)}{\partial y_t} z_t
- \frac{\partial g_s(y^*)}{\partial y_{t(s, \lambda)}} \sum_{t \in T_s, t \neq t(s, \lambda)} z_t \nonumber \\
= & \sum_{t \in T_s, t \neq t(s, \lambda)} \Bigl( \frac{\partial g_s(y^*)}{\partial y_t}
- \frac{\partial g_s(y^*)}{\partial y_{t(s, \lambda)}} \Bigr) z_t \nonumber \\
\leq & \quad 0. \nonumber
\end{align}

Thus, $y^*$ is an optimal solution to (\ref{formula:lagrangian_maxmization_2}).
Although $y^*$ may not be unique, $x_s^* = \sum_{t \in T_s} y^*_t$
is unique since $U_s(x_s)$ is strictly concave under assumption $A1$.

\noindent $(c)$
According to part $(b)$, for any $\lambda^* \in \Lambda^*$,
the solution $x^*$ obtained by (\ref{eq:sub_x})
is the unique Lagrangian maximizer with the optimal Lagrangian
multiplier $\lambda^*$. Furthermore, from part $(a)$, there is no duality gap. Thus, $x^*$
is optimal to (\ref{formula:primal}).
Since $U_s(x_s)$ is strictly concave in $x_s$ for any source $s$, $\sum_{s \in S} U_s(x_s)$
is strictly concave in the vector $x$. Hence, $x^*$ is the unique
optimal solution to (\ref{formula:primal}).

\noindent $(d)$
From part $(a)$, $\Lambda^*$ is non-empty.
Suppose $\Lambda^*$ is not bounded. We can make
$\theta (\lambda^*)$ arbitrarily large by choosing
$\lambda^* \in \Lambda^*$ with a large enough norm $||\lambda^*||$,
since (\ref{ineq:capacity_constr}) holds with strict inequality
at $(\bar{x}, \bar{y})$ under assumption $A2$.
This contradicts with the facts that $\theta(\lambda^*) = f^*$ and $f^*$ is bounded. Next,
since the objective function $f(x, y)$ is concave, it is continuous. $\Lambda^*$ is the inverse image
of the set $\{f^*\}$, and hence, is a closed set.
Therefore,
$\Lambda^*$ is a non-empty compact set.

\subsection{Proof of Theorem \ref{theorem:convergence_sub_lambda_x}}
\noindent {\bf Step size rule I}:
For any $\lambda^* \geq 0, \lambda^* \in \Lambda^*$,
by (\ref{eq:sub_lambda}) and by the non-expansive property of projection \cite{Bert99}, we have
\begin{align}
\label{ineq:theo_conv_proof}
& || \lambda(k + 1) - \lambda^* ||^2 \nonumber \\
\leq & || \lambda(k) - \delta (c - H y(k)) - \lambda^* ||^2 \nonumber \\
= & || \lambda(k) - \lambda^* ||^2 + \delta^2 || c - H y(k) ||^2 \nonumber \\
&    - 2 \delta (\lambda(k) - \lambda^*)^T (c - H y(k)).
\end{align}
By using the subgradient inequality \cite{Bert99},
\be
\theta(\lambda^*) \geq \theta(\lambda(k)) + (c - H y(k))^T (\lambda^* - \lambda(k)), \nonumber
\ee
we have
\be
\label{ineq:subgradient_feature}
-(\lambda(k) - \lambda^*)^T (c - H y(k)) \leq \theta(\lambda^*) - \theta(\lambda(k)).
\ee
Substituting (\ref{ineq:subgradient_feature})
into (\ref{ineq:theo_conv_proof}), we have
\begin{align}
\label{ineq:successive_lambda_dist}
& || \lambda(k + 1) - \lambda^* ||^2  \\
\leq & || \lambda(k) - \lambda^* ||^2 + \delta^2 || c - H y(k) ||^2
    + 2 \delta (\theta(\lambda^*) - \theta(\lambda(k))). \nonumber
\end{align}
Fix $\eta > 0$. Let
$\Lambda(\eta) = \{ \lambda \geq 0 \ | \ \theta(\lambda) \leq \theta(\lambda^*) + \eta\}$.
Since both $c$ and $x(k)$, and hence $y(k)$, are bounded,
there exists $M < \infty$ such that for all time $k$
\be
|| c - H y(k) ||^2 \leq M. \nonumber
\ee
If we pick $\delta \leq \frac{\eta}{M}$,
then as long as $\lambda(k) \not\in \Lambda(\eta)$,
i.e., $\theta(\lambda(k)) - \theta(\lambda^*) > \eta$,
from (\ref{ineq:successive_lambda_dist}), we have
\begin{align}
& || \lambda(k + 1) - \lambda^* ||^2 \nonumber \\
\leq & || \lambda(k) - \lambda^* ||^2 + \delta^2 M - 2 \delta \eta \nonumber \\
\leq & || \lambda(k) - \lambda^* ||^2 + \delta \frac{\eta}{M} M - 2 \delta \eta \nonumber \\
= & || \lambda(k) - \lambda^* ||^2 - \delta \eta. \nonumber
\end{align}
Hence, eventually, $\lambda(k)$ will enter the set $\Lambda(\eta)$. On
the other hand, if we pick $\delta \leq \frac{\eta}{\sqrt{M}}$, then
once $\lambda(k) \in \Lambda(\eta)$, we have
\bea
&& ||\lambda(k + 1) - \lambda^*|| \nonumber \\
& = & || [\lambda(k) - \delta (c - H y(k))]_+ - \lambda^* || \nonumber \\
& \leq & || \lambda(k) - \delta (c - H y(k)) - \lambda^* || \nonumber \\
& \leq & || \lambda(k) - \lambda^* || + \delta ||c - H y(k)|| \nonumber \\
& \leq & || \lambda(k) - \lambda^* || + \eta, \nonumber
\eea
where, again, the first equality is due to (\ref{eq:sub_lambda}), and the second inequality
holds due to the non-expansive property of projection. The third inequality is
due to the triangle inequality, and the last inequality holds by plugging in
the upper bounds of $\delta$ and $||c - H y(k)||$.

Since the above inequality holds for any $\lambda^* \in \Lambda^* \subseteq \Lambda(\eta)$,
it implies that
\be
d(\lambda(k + 1), \Lambda^*) \leq d(\lambda(k), \Lambda^*) + \eta, \nonumber
\ee
where $d(\lambda, \Lambda^*) = \min_{\lambda^* \in \Lambda^*} ||\lambda - \lambda^*||$.
Hence, if $\delta \leq \min \{\frac{\eta}{M}, \frac{\eta}{\sqrt{M}} \}$,
then there exists a time $K_0$ such that
$d(\lambda(k), \Lambda^*) \leq \xi(\eta)$ for all $k \geq K_0$,
where $\xi(\eta) = \max_{\lambda \in \Lambda(\eta)} d(\lambda, \Lambda^*) + \eta$.
It is easy to show that, as $\eta \rightarrow 0$,
$\xi(\eta) \rightarrow 0$.

Hence, for any $\epsilon > 0$, we can pick $\eta$ sufficiently
small such that $\xi(\eta) < \epsilon$. Then, there exists some
$\delta_0 = \min \{\frac{\eta}{M}, \frac{\eta}{\sqrt{M}} \} > 0$,
such that for any $\delta \leq \delta_0$ and any initial $\lambda(0) \geq 0$,
there exists a time $K_0$ such that
$d(\lambda(k), \Lambda^*) < \epsilon$ for all $k \geq K_0$.

\noindent {\bf Step size rule II}:
$d(\lambda(t), \Lambda^*) \rightarrow 0$ follows from \cite{BazaraaSS}.

Finally, since the mapping from $\lambda(k)$ to $x(k)$ is continuous,
we can pick $\eta$ sufficiently small and
there exists some $\delta_0 > 0$,
such that for any $\delta \leq \delta_0$,
$||x(k) - x^*|| < \epsilon$ for all $k \geq K_0$.

\subsection{Proof of Theorem \ref{theorem:convergence_sub_y}}
First, by Theorem \ref{theorem:convergence_sub_lambda_x}, the sequence
$\{ \lambda(k) \}$ converges to the compact set $\Lambda^*$ under
both step size rule I and II. Hence,
there exists a large enough constant $0 < \Delta_e < \infty$
such that $\lambda_e(k) \leq \Delta_e$ for all time $k$.
Next, according to (\ref{eq:sub_lambda}),
\be
\frac{1}{\delta_e(k)} (\lambda_e(k + 1) - \lambda_e(k)) \geq
    \sum_{t \in T: e \in t} y_t(k) - c_e, \forall e \in E. \nonumber
\ee
Summing the above inequality from time slots $k_0$ to $k$, we have
\begin{align}
& \sum_{u = k_0}^k \sum_{t \in T: e \in t} y_t(u) \nonumber \\
\leq & c_e (k - k_0 + 1)
+ \frac{1}{\delta_e(k)} \lambda_e(k + 1)
+ \sum_{u = k_0 + 1}^k (\frac{1}{\delta_e(u - 1)} \nonumber \\
& - \frac{1}{\delta_e(u)}) \lambda_e(u)
- \frac{1}{\delta_e(k_0)} \lambda_e(k_0)
\nonumber \\
\leq & c_e (k - k_0 + 1)
+ \frac{1}{\delta_e(k)} \Delta_e
+ \sum_{u = k_0 + 1}^k (\frac{1}{\delta_e(u - 1)} - \frac{1}{\delta_e(u)})
\Delta_e \nonumber \\
= & c_e (k - k_0 + 1) + \frac{1}{\delta_e(k_0)} \Delta_e \nonumber \\
= & c_e (k - k_0 + 1) + M_e. \nonumber
\end{align}
Note that $\frac{1}{\delta_e(u - 1)} - \frac{1}{\delta_e(u)} \geq 0$
under both step size rule I and II.
The last equality holds since ${\delta_e(k_0)} \neq 0$. Here, $M_e$ is
set to be $\Delta_e / \delta_e(k_0)$.

\subsection{Proof of Theorem \ref{theorem:avg_y_feasible}}
By Theorem \ref{theorem:convergence_sub_y}, for any link $e \in E$,
there exists a constant $M_e < \infty$ such that
\be
\label{ineq:converg_y_ineq}
\sum_{t \in T: e \in t} \bar{y}_t (k)
\leq c_e + \frac{M_e}{k - k_0 + 1}.
\nonumber
\ee

Taking the limits of the above inequality on both sides, it yields
\be
\lim_{k \rightarrow \infty} \sup
\sum_{t \in T: e \in t} \bar{y}_t (k) \leq
\lim_{k \rightarrow \infty} (c_e + \frac{M_e}{k - k_0 + 1}) = c_e,
\nonumber
\ee
for any link $e \in E$. Equivalently,
\be
\lim_{k \rightarrow \infty} \sup H \bar{y}(k) \leq c.
\nonumber
\ee

The sequence $\{ \bar{y} (k) \}$ is a bounded sequence since
$x(k)$ and $y(k)$ are bounded.
Hence, $\{ \bar{y} (k) \}$ has at least one limit point.
For any limit point $\bar{y}^*$ of the sequence $\{ \bar{y} (k) \}$,
there exists a subsequence $\{ \bar{y} (k) \}_{\mathcal{K}}$ converging
to $\bar{y}^*$, i.e.,
$\lim_{k \rightarrow \infty} \{ \bar{y} (k) \}_{\mathcal{K}}
= \bar{y}^*$.
By Theorem \ref{theorem:convergence_sub_y},
\be
H \bar{y}(k) \leq c + \frac{M}{k - k_0 + 1},  \forall k \in \mathcal{K}.
\nonumber
\ee
Since the subsequence $\{ H \bar{y} (k) \}_{\mathcal{K}}$ converges
as well by the continuity of the mapping from $y$ to $Hy$, we take the limits
on the both sides of the above inequality, which yields
\be
\lim_{k \rightarrow \infty, k \in \mathcal{K}} H \bar{y}(k)
\leq \lim_{k \rightarrow \infty, k \in \mathcal{K}} (c + \frac{M}{k - k_0 + 1})
= c.
\nonumber
\ee
By the continuity of the mapping from $y$ to $Hy$, we have
\be
H \bar{y}^*
= H \lim_{k \rightarrow \infty, k \in \mathcal{K}} \bar{y}(k)
= \lim_{k \rightarrow \infty, k \in \mathcal{K}} H \bar{y}(k)
\leq c.
\nonumber
\ee



\subsection{Proof of Theorem \ref{theorem:convergence_avg_y}}
Let $\bar{y}^*$ be a limit point of the sequence $\{ \bar{y}(k)\}$.
By Theorem \ref{theorem:avg_y_feasible}, we have
\be
\label{ineq:converg_y_lp_1}
H \bar{y}^* \leq c.
\ee

At any time slot $k$, $A y(k) = x(k)$ by (\ref{eq:sub_y}).
By Theorem \ref{theorem:convergence_sub_lambda_x},
for any $\epsilon > 0$, there exists a sequence of step size
$\{ \delta(k) \}$ and a sufficiently large $K_0$ such that,
for any initial $\lambda(0) \geq 0$, for all $k \geq K_0$,
$d(\lambda(k), \Lambda^*) < \epsilon$ and $||x(k) - x^*|| < \epsilon$.
Hence, for all $k \geq K_0$, $||A y(k) - x^*|| \leq \epsilon$.
It is easy to see that there exists a time $K_1 \geq K_0$ such that, for all
$k \geq K_1$, $||A \bar{y}(k) - x^*|| \leq \epsilon$.
Hence,
\be
\label{ineq:converg_y_lp_2}
||A \bar{y}^* - x^*|| \leq \epsilon.
\ee
From (\ref{ineq:converg_y_lp_1}) and (\ref{ineq:converg_y_lp_2}),
we have $\bar{y}^* \in \mathcal{Y}^* (\epsilon)$.

\section{Appendix B: Proofs in Section \ref{sec:colgen}}

\subsection{Proof of Lemma \ref{fact:RMP_opt_cond}}
Since the strong duality holds for both the master and the restricted problems,
we have
\be
\label{eqs:MP_RMP_strong_duality}
\sum_{s \in S} U_s (x_s^*) = \theta(\lambda^*), \ \ \
\sum_{s \in S} U_s (\bar{x}_s^{(q)}) = \theta^{(q)}(\bar{\lambda}^{(q)}).
\ee
Since the
$q^{th}$-RMP is more restricted than the MP, we have
\be
\label{eq:obj_lb_1}
\sum_{s \in S} U_s(x_s^*) \geq \sum_{s \in S} U_s(\bar{x}_s^{(q)}).
\ee
Combining (\ref{eqs:MP_RMP_strong_duality}) and (\ref{eq:obj_lb_1}),
we get the following lower bound for the optimal objective
value of the MP.
\be
\label{eq:obj_lb_2}
\sum_{s \in S} U_s(x_s^*)
\geq \sum_{s \in S} U_s(\bar{x}_s^{(q)})
= \theta^{(q)}(\bar{\lambda}^{(q)}).
\ee

By the weak duality \cite{Bert99},
for any $\lambda$ feasible to the dual problem of the MP,
$\theta(\lambda)$ is an upper bound for the optimal
objective value of the MP. In particular,
consider $\bar{\lambda}^{(q)}$, which is
optimal to the dual of the $q^{th}$-RMP and feasible to the dual of
the MP. $\theta (\bar{\lambda}^{(q)})$ is an upper bound of
$\sum_{s \in S} U_s(x_s^*)$, i.e.,
\be
\label{eq:obj_ub}
\theta(\bar{\lambda}^{(q)}) \geq \sum_{s \in S} U_s(x_s^*).
\ee

By the definitions of the dual functions,
\begin{align}
\label{eq:dual_obj_value_MP}
& \theta (\bar{\lambda}^{(q)}) -  \theta^{(q)} (\bar{\lambda}^{(q)})\nonumber \\
=& \sum_{s \in S} \Bigl( \max_{\substack{x_s = \sum_{t \in T_s} y_t,\\
    \ m_s \leq x_s \leq M_s, \ y \geq 0}}
    \{ U_s(x_s) -
\sum_{t \in T_s} y_t \sum_{e \in t} \bar{\lambda}^{(q)}_e \} \nonumber \\
& \quad - \max_{\substack{x_s = \sum_{t \in T_s^{(q)}} y_t, \\ \ m_s \leq x_s \leq M_s, \ y \geq 0}}
\{ U_s(x_s) -
\sum_{t \in T_s^{(q)}} y_t \sum_{e \in t} \bar{\lambda}^{(q)}_e \} \Bigr) \nonumber \\
=& \sum_{s \in S}
(h_s(\gamma(s, \bar{\lambda}^{(q)})) - h_s(\gamma^{(q)} (s, \bar{\lambda}^{(q)}))).
\nonumber
\end{align}
In the last equality, we plug in the Lagrangian maximizers according
to Lemma \ref{lemma:sub} part $(b)$.
Hence, the gap between the upper and lower bounds for the optimal objective
value of the MP is
$\sum_{s \in S}
(h_s(\gamma(s, \bar{\lambda}^{(q)})) - h_s(\gamma^{(q)} (s, \bar{\lambda}^{(q)})))$.
We will show later in the proof of Theorem \ref{theo:colgen_bound} that
$h_s(w)$ is a non-increasing function. Thus,
$\gamma(s, \bar{\lambda}^{(q)}) \leq \gamma^{(q)} (s, \bar{\lambda}^{(q)})$
implies $h_s(\gamma(s, \bar{\lambda}^{(q)})) - h_s(\gamma^{(q)} (s, \bar{\lambda}^{(q)})) \geq 0$
for any $s \in S$. Then, the optimality gap is $0$
if and only if
$h_s(\gamma(s, \bar{\lambda}^{(q)})) = h_s(\gamma^{(q)} (s, \bar{\lambda}^{(q)}))$
for all source $s \in S$.

\subsection{Proof of Theorem \ref{theo:convergence_colgen}}
Since the number of trees
in $T$ is finite, eventually Algorithm
\ref{algo:colgen_imperfect} will stop introducing new trees.
Hence, there exists a $q$, $1 \leq q \leq |T|$, such that,
after Algorithm \ref{algo:colgen_imperfect} stops introducing
new trees, the number of trees
that have been introduced is $q$. Let the subset containing
these $q$ trees be denoted by $T^{(q)}$.
After Algorithm \ref{algo:colgen_imperfect} no longer introduces
new trees, it behaves
just like the subgradient algorithm but on the restricted set
$T^{(q)}$.
According to the theorems in Section \ref{sec:subgradient}, the
subgradient algorithm converges.
Thus, Algorithm \ref{algo:colgen_imperfect} converges
to $(\bar{x}^{(q)}, \bar{\lambda}^{(q)})$
on this particular $q^{th}$-RMP.

We next show that, after Algorithm \ref{algo:colgen_imperfect} converges to
$(\bar{x}^{(q)}, \bar{\lambda}^{(q)})$, we have
$\gamma_{\rho} (s, \bar{\lambda}^{(q)}) = \gamma^{(q)} (s, \bar{\lambda}^{(q)})$
for any source $s \in S$.
First, note that
$\gamma_{\rho} (s, \bar{\lambda}^{(q)}) \leq \gamma^{(q)} (s, \bar{\lambda}^{(q)})$
by the comment after Algorithm \ref{algo:colgen_imperfect}. Next, it must
be true that $\gamma_{\rho} (s, \bar{\lambda}^{(q)}) \geq \gamma^{(q)} (s, \bar{\lambda}^{(q)})$.
Otherwise, the tree whose cost
is $\gamma_{\rho} (s, \bar{\lambda}^{(q)})$ must not have already been in
$T^{(q)}$ and will be selected to enter. This violates
the assumption that the algorithm never selects more than $q$ trees.

\subsection{Proof of Theorem \ref{theo:colgen_bound}}

Recall that we define $h_s(w)$ for each source $s$ as
\be
h_s(w) = U_s([(U'_s)^{-1}(w)]_{m_s}^{M_s})
    - [(U'_s)^{-1}(w)]_{m_s}^{M_s} \cdot w \nonumber
\ee
for all $w > 0$.
Let us first prove the monotonicity of $h_s(w)$.
\begin{lemma}
\label{lemma:h_monotonicity}
For any source $s$, $h_s(w)$ is a non-increasing function
for all $w > 0$, i.e.,
\be
\label{ineq:h_decreasing}
h_s(w_1) \geq h_s(w_2) \mbox{ if } 0 < w_1 \leq w_2.
\ee
\end{lemma}
\IEEEproof{
For each source $s$, we define a function
\be
g_s(w) = U_s((U'_s)^{-1}(w)) - (U'_s)^{-1}(w) \cdot w \nonumber
\ee
for all $w > 0$. $g_s(w)$ is a non-increasing function for
all $w > 0$. This monotonicity can be verified by checking
$g'_s(w)$.
\begin{align}
g'(w)
= & U'_s((U'_s)^{-1}(w)) \cdot ((U'_s)^{-1})'(w) \nonumber \\
& \quad - ((U'_s)^{-1})'(w) \cdot w - (U'_s)^{-1}(w) \nonumber \\
= & w \cdot ((U'_s)^{-1})'(w)
  - ((U'_s)^{-1})'(w) \cdot w - (U'_s)^{-1}(w) \nonumber \\
= & - (U'_s)^{-1}(w) \leq 0. \nonumber
\end{align}
Here, $U_s(\cdot)$ is a non-decreasing function and
$U'_s(x) \geq 0$ for all $0 \leq m_s \leq x \leq M_s$.
Hence $(U'_s)^{-1}(w) \geq 0$ for all $w > 0$.

We also note that $U'_s(\cdot)$ is a decreasing function
since $U_s(\cdot)$ is strictly concave.
Hence $(U'_s)^{-1}(\cdot)$ is also a decreasing function.
To prove the monotonicity of the function $h_s(w)$, we need
to discuss serval cases.

\noindent {\bf Case $1$}:
$U'_s(M_s) \leq w_1 \leq w_2 \leq U'_s(m_s)$.
From the monotonicity of $(U'_s)^{-1}(\cdot)$,
$[(U'_s)^{-1}(w_1)]_{m_s}^{M_s} = (U'_s)^{-1}(w_1)$,
and $[(U'_s)^{-1}(w_2)]_{m_s}^{M_s} = (U'_s)^{-1}(w_2)$
in this case.
From $0 < w_1 \leq w_2$ and the fact that
$g_s(w)$ is a non-increasing function, we have
$g_s(w_1) \geq g_s(w_2)$, which yields
(\ref{ineq:h_decreasing}).

\noindent {\bf Case $2$}:
$w_1 \leq U'_s(M_s) \leq w_2 \leq U'_s(m_s)$.
As in case $1$, we can show that
\begin{align}
& U_s([(U'_s)^{-1}(w_2)]_{m_s}^{M_s}) -
[(U'_s)^{-1}(w_2)]_{m_s}^{M_s} \cdot w_2
\leq \nonumber \\
& U_s([(U'_s)^{-1}(U'_s(M_s))]_{m_s}^{M_s}) -
[(U'_s)^{-1}(U'_s(M_s))]_{m_s}^{M_s} \cdot U'_s(M_s). \nonumber
\end{align}
Furthermore,
\begin{align}
& U_s([(U'_s)^{-1}(w_1)]_{m_s}^{M_s}) -
[(U'_s)^{-1}(w_1)]_{m_s}^{M_s} \cdot w_1 \nonumber \\
= & U_s(M_s) - M_s \cdot w_1 \nonumber \\
\geq & U_s(M_s) - M_s \cdot U'_s(M_s) \nonumber \\
= & U_s([(U'_s)^{-1}(U'_s(M_s))]_{m_s}^{M_s}) -
[(U'_s)^{-1}(U'_s(M_s))]_{m_s}^{M_s} \cdot U'_s(M_s). \nonumber
\end{align}
Hence, (\ref{ineq:h_decreasing}) holds by combining the above
two inequalities.

\noindent {\bf Case $3$}:
$w_1 \leq w_2 \leq U'_s(M_s)$.
In this case
\be
U_s([(U'_s)^{-1}(w_2)]_{m_s}^{M_s}) -
[(U'_s)^{-1}(w_2)]_{m_s}^{M_s} \cdot w_2
= U_s(M_s) - M_s \cdot w_2, \nonumber
\ee
and
\be
U_s([(U'_s)^{-1}(w_1)]_{m_s}^{M_s}) -
[(U'_s)^{-1}(w_1)]_{m_s}^{M_s} \cdot w_1
= U_s(M_s) - M_s \cdot w_1. \nonumber
\ee
Hence, (\ref{ineq:h_decreasing}) holds.

\noindent {\bf Case $4$}:
$U'_s(M_s) \leq w_1 \leq U'_s(m_s) \leq w_2$.
The proof is the same as that of case $2$.

\noindent {\bf Case $5$}:
$U'_s(m_s) \leq w_1 \leq w_2$.
(\ref{ineq:h_decreasing}) holds by a similar argument as in case $3$.
}

With Lemma \ref{lemma:h_monotonicity}, we proceed to
the proof of Theorem \ref{theo:colgen_bound}.
Since the $q^{th}$-RMP is more restricted than the MP, we have
$\theta^{(q)} (\bar{\lambda}^{(q)})
\leq \sum_{s \in S} U_s(x_s^*)$.
Note that $\rho \bar{\lambda}^{(q)} \geq 0$
is a feasible dual variable.
By the weak duality, we have
$\sum_{s \in S} U_s(x_s^*)
\leq \theta (\rho \bar{\lambda}^{(q)})$.

By the definition of the dual functions, we have
\begin{align}
 \theta(\rho \bar{\lambda}^{(q)})
= & \sum_{s \in S} \max_{\substack{x_s = \sum_{t \in T_s} y_t,\\
    \ m_s \leq x_s \leq M_s, \ y \geq 0}} \bigl( U_s(x_s) \nonumber \\
  & \quad - \sum_{t \in T_s} y_t \sum_{e \in t} \rho \bar{\lambda}^{(q)}_e \bigr)
    + \sum_{e \in E} \rho \bar{\lambda}^{(q)}_e c_e
    \nonumber \\
= & \sum_{s \in S} h_s(\rho \gamma (s, \bar{\lambda}^{(q)}))
    + \sum_{e \in E} \rho \bar{\lambda}^{(q)}_e c_e.
    \nonumber
\end{align}
To see the second equality, we note that
after the dual variable is linearly scaled up by $\rho$,
the global min-cost tree is not changed and
the global minimum tree cost is
$\rho \gamma (s, \bar{\lambda}^{(q)})$.
By similar arguments
as in the proof of Theorem \ref{lemma:sub},
we recognize that
one of the Lagrangian maximizers
of $\theta(\cdot)$ at $\rho \bar{\lambda}^{(q)}$
is
\be
x_s(\rho \bar{\lambda}^{(q)})
= [(U'_s)^{-1} (\rho \gamma (s, \bar{\lambda}^{(q)}))]_{m_s}^{M_s},
\forall s \in S, \nonumber
\ee
and
\be
y_t(\rho \bar{\lambda}^{(q)}) =
\left\{ \begin{array}{ll}
         x_s(\rho \bar{\lambda}^{(q)}) & \mbox{if
         $t = t(s, \rho \bar{\lambda}^{(q)})$ for some $s \in S$;}\\
         0 & \mbox{otherwise}. \end{array} \right.
  \nonumber
\ee

Without loss of generality, we can assume that
$\gamma (s, \bar{\lambda}^{(q)}) > 0$.
From (\ref{ineq:tree_approx_ratio}), we have
$0 < \gamma (s, \bar{\lambda}^{(q)})
\leq \gamma_{\rho} (s, \bar{\lambda}^{(q)})
\leq \rho \gamma (s, \bar{\lambda}^{(q)})$,
which implies
\be
\label{ineq:lagrangian_fun_compare}
h_s(\rho \gamma (s, \bar{\lambda}^{(q)}))
\leq h_s(\gamma_{\rho} (s, \bar{\lambda}^{(q)})).
\ee

Now the dual function $\theta(\rho \bar{\lambda}^{(q)})$
can be written explicitly as
\begin{align}
\theta(\rho \bar{\lambda}^{(q)})
=& \rho \sum_{e \in E} \bar{\lambda}^{(q)}_e c_e
    + \sum_{s \in S} h_s(\rho \gamma (s, \bar{\lambda}^{(q)}))
    \nonumber \\
\leq & \rho \sum_{e \in E} \bar{\lambda}^{(q)}_e c_e
    + \sum_{s \in S} h_s(\gamma_{\rho} (s, \bar{\lambda}^{(q)}))
    \nonumber \\
= & \rho \sum_{e \in E} \bar{\lambda}^{(q)}_e c_e
    + \sum_{s \in S} h_s(\gamma^{(q)} (s, \bar{\lambda}^{(q)}))
    \nonumber \\
\leq & \rho \sum_{e \in E} \bar{\lambda}^{(q)}_e c_e
    + \rho \sum_{s \in S} h_s(\gamma^{(q)} (s, \bar{\lambda}^{(q)}))
    \nonumber \\
= & \rho \theta^{(q)} (\bar{\lambda}^{(q)}). \nonumber
\end{align}
The first equality is by plugging in the Lagrangian maximizer at
$\rho \bar{\lambda}^{(q)}$.
The second inequality is from (\ref{ineq:lagrangian_fun_compare}).
The third equality holds because
$\gamma_{\rho} (s, \bar{\lambda}^{(q)}) =
\gamma^{(q)} (s, \bar{\lambda}^{(q)})$ by
Theorem \ref{theo:convergence_colgen}. The fourth inequality holds
because $\rho \geq 1$ and, under the assumptions $A3$ and $A4$,
$h_s(\gamma^{(q)} (s, \bar{\lambda}^{(q)})) \geq 0$.
To see that $h_s(\gamma^{(q)} (s, \bar{\lambda}^{(q)})) \geq 0$,
we note that $\gamma^{(q)} (s, \bar{\lambda}^{(q)}) < U'_s(m_s)$
by the assumption $A3$ (i.e., $\bar{x}_s^{(q)} > m_s$).
Since $h_s(w)$ is non-increasing, we have
$h_s(\gamma^{(q)} (s, \bar{\lambda}^{(q)}))
\geq h_s(U'_s(m_s)) = U_s(m_s) - m_s \cdot U'_s(m_s) \geq 0$.
The non-negativity of $h_s(U'_s(m_s))$ is by the assumption
$A4$. The last equality holds by recognizing that
$\theta^{(q)} (\bar{\lambda}^{(q)})
= \sum_{e \in E} \bar{\lambda}^{(q)}_e c_e
    + \sum_{s \in S} h_s(\gamma^{(q)} (s, \bar{\lambda}^{(q)}))$.

\bibliographystyle{IEEEtran}
\bibliography{MultiMulticast}

\begin{IEEEbiography}
{Xiaoying Zheng}
received the bachelor's and master's degrees in computer science and engineering
from Zhejiang University, P.R. China, in 2000 and 2003, respectively, and the
PhD degree in computer engineering from the University of Florida, Gainesville, in 2008.
She was an assistant professor in Shanghai Research Center of Wireless Communications, from 2009 to 2011. She then joined Shanghai Advanced Research Institute, Chinese Academy of Sciences in 2012 as an associate professor. Her research interests include applications of optimization theory in networks, distributed systems, performance evaluation of network protocols and algorithms, peer-to-peer overlay networks, content distribution, and congestion control.
\end{IEEEbiography}

\begin{IEEEbiography}
{Chunglae Cho}
received the B.S. and M.S. degrees in computer science from Pusan National University, Korea, in 1994 and 1996, respectively. He worked as a research staff member at Electronics and Telecommunications Research Institute, Korea, between 2000 and 2005. He is currently working toward a Ph.D. degree at the Computer and Information Science and Engineering department at the University of Florida, Gainesville, FL. His research interests are in computer networks, including resource allocation and optimization on peer-to-peer networks and wireless networks.
\end{IEEEbiography}

\begin{IEEEbiography}
{Ye Xia} is an associate professor at the Computer and Information
Science and Engineering department at the University of Florida,
starting in August 2003. He has a PhD degree from the University
of California, Berkeley, in 2003, an MS degree in 1995 from
Columbia University, and a BA degree in 1993 from Harvard
University, all in Electrical Engineering. Between June 1994 and
August 1996, he was a member of the technical staff at Bell
Laboratories, Lucent Technologies in New Jersey. His main research
area is computer networking, including performance evaluation of
network protocols and algorithms, resource allocation, wireless
network scheduling, network optimization, and load balancing on
peer-to-peer networks. He also works on cache organization and
performance evaluation for chip multiprocessors. He is interested
in applying probabilistic models to the study of computer systems.
\end{IEEEbiography}

\end{document}